\documentclass[twocolumn]{aastex62}
\usepackage{amsmath}

\graphicspath{{./}{}}

\newcommand{\EQ}{\begin{equation}}
\newcommand{\EE}{\end{equation}}
\newcommand{\EQA}{\begin{eqnarray}}
\newcommand{\EEA}{\end{eqnarray}}

\newcommand\T{\rule{0pt}{2.0ex}}       
\newcommand{\mso}{\,\mathrm{M}_\odot}
\newcommand{\rso}{\,{\rm R}_\odot}
\newcommand{\lso}{\,{\rm L}_\odot}
\newcommand{\hrd}{HR diagram}
\newcommand{\kms}{\, {\rm km}\, {\rm s}^{-1}}

\newcommand{\teff}{\log\, T_{\rm eff}\,}

\newcommand{\vca}{v_{\mathrm {c}}}
\newcommand{\hp}{H_{\mathrm{P}}}

\newcommand{\grad}{\nabla}
\newcommand{\adrad}{\nabla_{\mathrm{\!rad}}}
\newcommand{\gradad}{\nabla_{\mathrm{\!ad}}}

\newcommand{\alphaov}{\ensuremath{\alpha_{\mathrm{ov}}}} 
\newcommand{\alphamlt}{\ensuremath{\alpha_{\mathrm{MLT}}}} 
\newcommand{\cp}{c_{\rm P}}

\shorttitle{Envelope Convection, Surface Magnetism and Spots in A and late B type stars}
\shortauthors{Cantiello \& Braithwaite}

\begin{document}
\title{ENVELOPE CONVECTION, SURFACE MAGNETISM AND SPOTS IN A AND LATE B-TYPE STARS}

\correspondingauthor{Matteo Cantiello}
\email{mcantiello@flatironinstitute.org}

\author{Matteo Cantiello}
\affiliation{Center for Computational Astrophysics, Flatiron Institute, 162 5th Avenue, New York, NY 10010, \
USA}
\affiliation{Department of Astrophysical Sciences, Princeton University, Princeton, NJ 08544, USA}

\author{Jonathan Braithwaite}
\affiliation{Argelander-Institut f\"ur Astronomie der Universit\"at Bonn, Auf dem H\"ugel 71, D--53121 Bonn, Germany}

\begin{abstract}

Weak magnetic fields have recently been detected in a number of A-type stars, including Vega and Sirius. At the same time, space photometry observations of A and late B-type stars from Kepler and TESS have highlighted the existence of rotational modulation of surface features akin to stellar spots. Here we explore the possibility that surface magnetic spots might be caused by the presence of small envelope convective layers at or just below the stellar surface, caused by recombination of H and He. Using 1D stellar evolution calculations and assuming an equipartition dynamo, we make simple estimates of field strength at the photosphere. For most models the largest effects are caused by a convective layer driven by second helium ionization. While it is difficult to predict the geometry of the magnetic field, we conclude that the majority of intermediate-mass stars 
should have dynamo-generated magnetic fields of order a few gauss at the surface.  These magnetic fields can appear at the surface as bright spots, and  cause photometric variability via rotational modulation, which could also be wide-spread in A-stars.  The amplitude of surface magnetic fields and their associated photometric variability is expected to decrease with increasing stellar mass and surface temperature, so that magnetic spots and their observational effects should be much harder to detect in late B-type stars.  
\end{abstract}


\section{Introduction}
To a first approximation, main-sequence stars above around 1.5$\mso$ have a convective core and a radiative envelope. A more careful inspection reveals the presence of thin convective layers at or just below the photosphere, caused by opacity bumps and a decrease of the adiabatic gradient associated with the ionization of hydrogen, helium and iron-group elements.

\citet{2009A&A...499..279C} investigated the subsurface iron-ionization-driven convective layer (FeCZ), which in solar metallicity stars appear above about $7\mso$. Beside being important for driving surface turbulence, these convective layers could also host a dynamo, from where there is no difficulty for the resulting magnetic field to reach the surface via buoyancy \citep{2011A&A...534A.140C}. Surface field strengths of (very approximately) $5$ to $300$ G are predicted, depending on the mass and age of the star, with higher fields in more massive stars and towards the end of the main sequence. The FeCZ should play a role in various observational effects such as line profile variability, discrete absorption components and photometric variability, amongst others \citep{2009A&A...499..279C}. In particular, the bright spots associated with emergent magnetic fields predicted by \citet{2011A&A...534A.140C} might have been recently observed by the MOST and BRITE missions \citep{2014MNRAS.441..910R,2018MNRAS.473.5532R}. Short-lived, co-rotating bright spots can explain part of the photometric variability in a large fraction of the observed OB targets, and therefore could be an ubiquitous feature of these stars. The same subsurface convection that can produce these spots could also be responsible for the  low-frequency power excess observed in a number of massive targets by both CoRoT and the BRITE constellation \citep[See e.g.][]{2011A&A...533A...4B,2018MNRAS.480..972R,2018arXiv181108023B}.
The question then arises: Can the less vigorous surface and subsurface regions associated with hydrogen and helium ionization in stars between $\simeq1.5$ and 7$\mso$ also produce observable effects? The possibility that these regions could lead to surface magnetism and spots is the subject of this paper.

Early-type stars display a bimodality in their magnetic properties (\citealt{2007AA4751053A}; see \citealt{2009ARA&A..47..333D} and \citealt{2017RSOS....460271B} for reviews of magnetic fields in relevant kinds of stars). The chemically-peculiar Ap and Bp stars (accounting for a few percent of the population) have large-scale magnetic fields with strengths ranging from about $200$ G to over $30$ kG. These fields have not been seen to change with time, over decades of observations. Similar magnetic fields have been found in more massive OB stars. These magnetic fields are thought to be ``fossils'': a magnetic field in stable equilibrium, left over from some earlier epoch \citep{1945MNRAS.105..166C,2004Natur.431..819B}. 
The Ap stars display unusually high abundances of rare earths and some lighter elements such as silicon, and  inhomogeneities of these elements on the surface show correlation with the magnetic field structure. This is thought to be connected to a magnetic suppression of turbulence and convection, allowing the separation of different species via radiative levitation and gravitational settling near the stellar surface \citep{1970ApJ...160..641M}. 
Amongst the other A stars, various other chemical peculiarities are seen, for instance in the slowly-rotating Am stars -- of which Sirius is the best-known specimen -- and mercury-manganese stars. See \citealt{2003ASPC..305..199T} for a review of these `skin diseases'.

Concerning the magnetic properties of the rest of the A-star population, one can generally only place upper limits of a few gauss on the large-scale magnetic fields \citep{2010A&A...523A..41P}, although it is possible in principle that a small-scale field somewhat stronger than that is present. 
In more massive stars the detection limit is rather higher than in A stars \citep[e.g.][]{2008A&A...483..857S}, but there is still a clear bimodality. For a few A-stars, ultra-deep Zeeman polarimetric observations have been performed. Surprisingly, these observations revealed the presence of ultra-weak magnetic fields: In Vega a field of $0.6\pm0.3$ G was found \citep{2009A&A...500L..41L,2010A&A...523A..41P}, and in the Am star Sirius a field of $0.2\pm0.1$ G has been measured \citep{2011A&A...532L..13P}. Recently a few more Am stars have been discovered to posses ultra-weak magnetic fields \citep{Blazere:2016b,Blazere:2016a}, possibly hinting to an ubiquitous nature of low-amplitude magnetism in these objects. 
Note that stars with magnetic fields with amplitude between 300 and a few G are not detected  \citep{2007AA4751053A}.

The star Vega also shows a spectroscopic variability revealing the presence of surface structures compatible with stable rotational modulation of hot or cold stellar spots \citep{Bohm:2015}, with some  features apparently stable on long timescales, and other evolving on timescales of days \citep{2017MNRAS.472L..30P}. 
The presence of spots in intermediate-mass stars might be wide-spread, with recent Kepler, K2 and TESS observations revealing that the most common variability type among mid-A to late-B stars is a simple periodic variability tentatively associated with rotational modulation of surface spots \citep{Balona:2017,Balona:2019}.

In a previous paper \citep{2013MNRAS.428.2789B} we introduced the idea that weak fields in A-type stars could be `failed fossils', fields which are evolving dynamically towards a `fossil' equilibrium, but have not yet arrived there. Since the field evolves dynamically on a timescale $\tau_{\rm evol}\sim\tau_{\rm A}^2\Omega$, where $\tau_{\rm A}$ and $\Omega$ are the Alfv\'en timescale and the rotational angular velocity, if the field is weak and the rotation fast, the field evolves very slowly. Equating the age of Vega and Sirius to this dynamical evolution timescale gives a field strength of order $10$ G.
Here we will discuss an alternative explanation, dynamo-generated magnetic fields originating in the subsurface convective layer driven by second helium ionization.

In this paper we investigate the presence and observational consequences of subsurface convective layers in intermediate-mass stars.  In Sect.~\ref{convection} we discuss the physics and occurrence of convection in the envelopes of intermediate stars. In Sect.~\ref{spots} the magnetic fields produced within the convective layers and their appearance at the surface is examined. We look at the observational consequences of these surface magnetic fields in Sect.~\ref{consequences}, and discuss the specific case of the star Vega in Sect.~\ref{sec:vega}. We summarize current ideas for the possible sources of magnetism in A stars in Sect.~\ref{magnetism_summary}. We discuss the results and conclude in Sects.~\ref{disc} and \ref{conc}.

\section{Occurrence of Convection}\label{convection}
According to the Schwarzschild criterion, convection occurs when the radiative gradient becomes larger than the adiabatic one. That is
\begin{equation}
\label{eqn:schwarzschild1}
\nabla_{\rm rad} = \frac{3F\kappa}{4acg}\frac{P}{T^4} > \nabla_{\rm ad},
\end{equation}
where $F$ is the flux density, $\kappa$ is the opacity and $\nabla_{\rm ad}\equiv (\partial \ln T/\partial\ln P)_{\rm ad}$ is the change in temperature in response to an adiabatic change in pressure. This comes from the equation of radiative transfer and the equation of hydrostatic equilibrium. Defining the following quantities
\begin{equation}
\nabla_F\equiv\frac{{\rm d}\ln F}{{\rm d}\ln P};\;\;\;\;\;\;\;\;
\nabla_g\equiv\frac{{\rm d}\ln g}{{\rm d}\ln P};\;\;\;\;\;\;\;\;
\nabla_\kappa\equiv\frac{{\rm d}\ln \kappa}{{\rm d}\ln P} 
\end{equation}
we can rewrite the condition for convection as
\begin{equation}
\label{eqn:schwarzschild2}
\frac{1}{4}(1+\nabla_F-\nabla_g+\nabla_\kappa) > \nabla_{\rm ad}.
\end{equation}
In intermediate-mass stars during the main sequence, radiation pressure is small compared to ideal-gas pressure and $\nabla_{\rm ad}\approx 2/5$ away from ionization zones, and there is no convection in the bulk of the star's volume. Both $\nabla_F$ and $\nabla_g$ are $-\infty$ in the very centre of the star, pass through zero at some radius and are positive beyond that radius. Now, because of the steep dependence on temperature of the reaction rate of the CNO cycle, much steeper than the p-p chain ($\epsilon_{\rm pp}\propto \rho T^{4}$, while $\epsilon_{\rm CNO}\propto \rho T^{18}$), nuclear heating is very centrally concentrated in early-type stars. Consequently, moving outwards from the centre $\nabla_F$ turns from negative to positive at a much smaller radius than does $\nabla_g$ and $\nabla_F-\nabla_g$ is positive in the core of the star, giving rise to convection.

\begin{figure}[h!]
\begin{center}
\includegraphics[width=1.00\columnwidth]{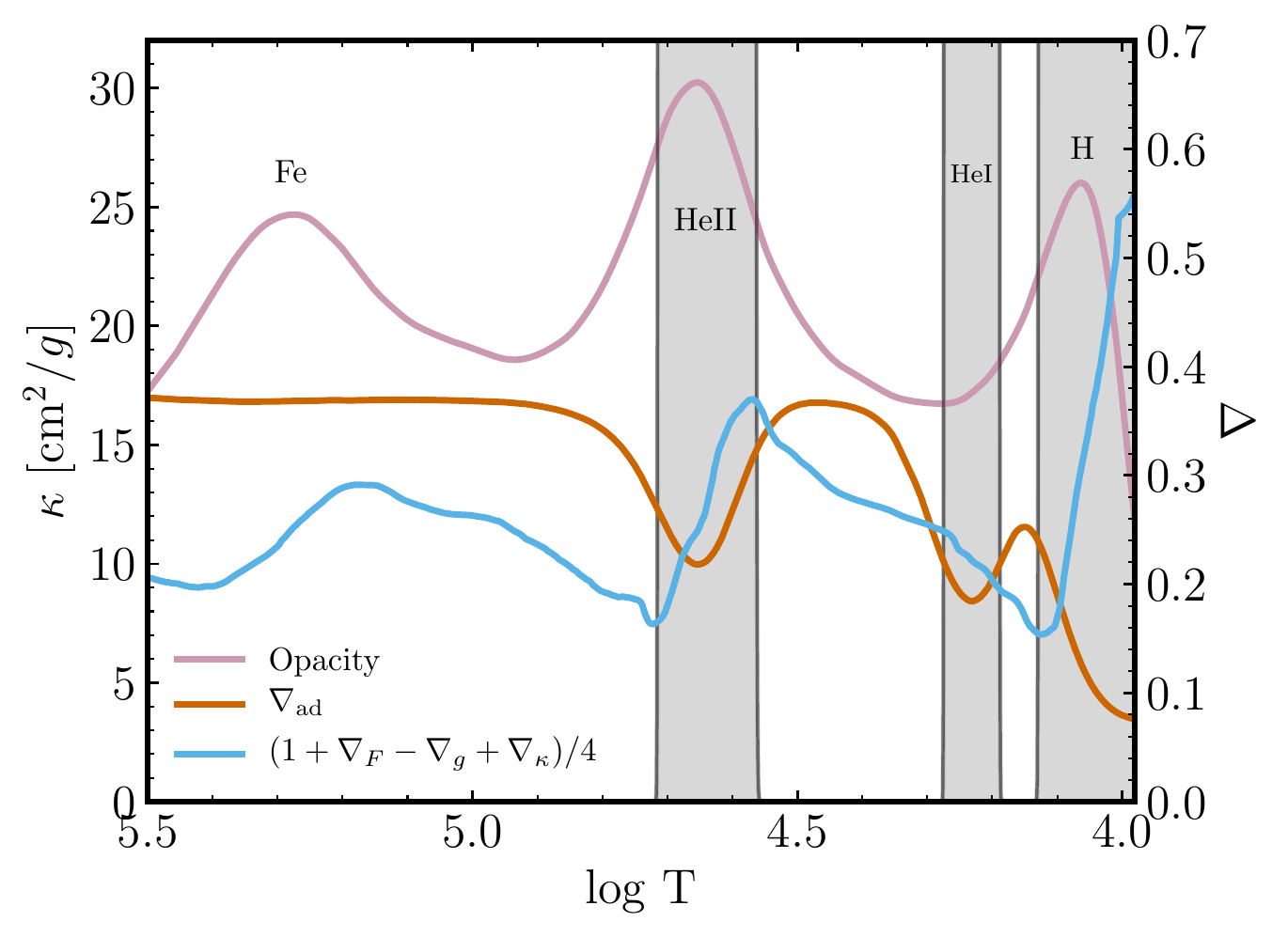}
\caption{{\label{Convection} Values of opacity $\kappa$ and adiabatic gradient $\nabla_{\rm ad}$ as function of temperature in the outer regions of a stellar model with mass 2.4$\mso$ and core hydrogen fraction X$_{c} = 0.47$. The stellar surface is located to the right. Convective regions are shaded grey, and we annotate the location of the  species undergoing partial ionization and driving convection. Note that, despite the presence of an opacity peak, the Fe convection zone is not present because of the insufficient luminosity of this model.  
}}
\end{center}
\end{figure} 

In contrast to this, near the stellar surface both $\nabla_F$ and $\nabla_g$ are small and cancel each other out almost perfectly since the outer layers of the star contain very little mass and produce no nuclear energy. For convection we need therefore either a large opacity gradient $\nabla_\kappa$ (i.e.\ opacity increasing inwards), and/or a high heat capacity and consequently low $\nabla_{\rm ad}$\footnote{Conceptually one can draw a parallel here with cumulonimbus clouds: convection caused by heating from below can rise to great heights above the cloud base, driven by the release of latent heat, analogous to heat from recombination of ionised species. Convection inside a cloud requires a temperature gradient (`lapse rate' in atmospheric parlance) of as little as $3^\circ$C/km, as opposed to $10^\circ$C/km in non-saturated air.}, see Fig.~\ref{Convection}. 
We explore how this happens in the following section.

\begin{figure*}[ht!]
\includegraphics[width=\textwidth]{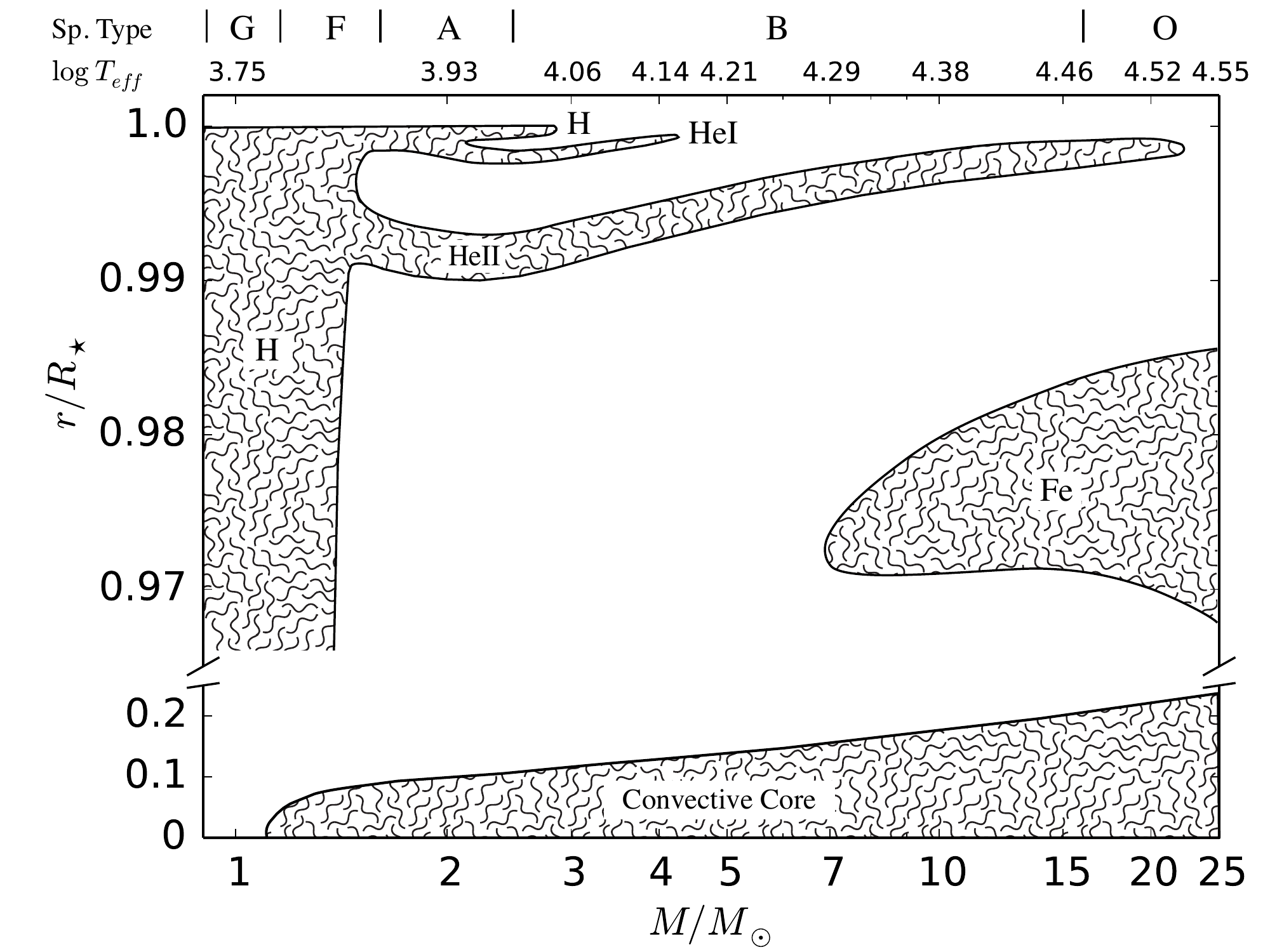}
\caption{{\label{Kipp}  Normalized radial extension of core, surface and subsurface convection zones for stars in the mass range 0.9-25 $\mso$. The models are extracted during the main sequence when the center mass fraction of H is 0.5. The convective regions are associated with ionization of H, He (HeI and HeII) and iron peak elements (Fe). The stellar surface $r/R_{*} = 1$ is defined as the location corresponding to optical depth $\tau=2/3$.%
}}
\end{figure*}

\begin{figure}[h!]
\begin{center}
\includegraphics[width=1.00\columnwidth]{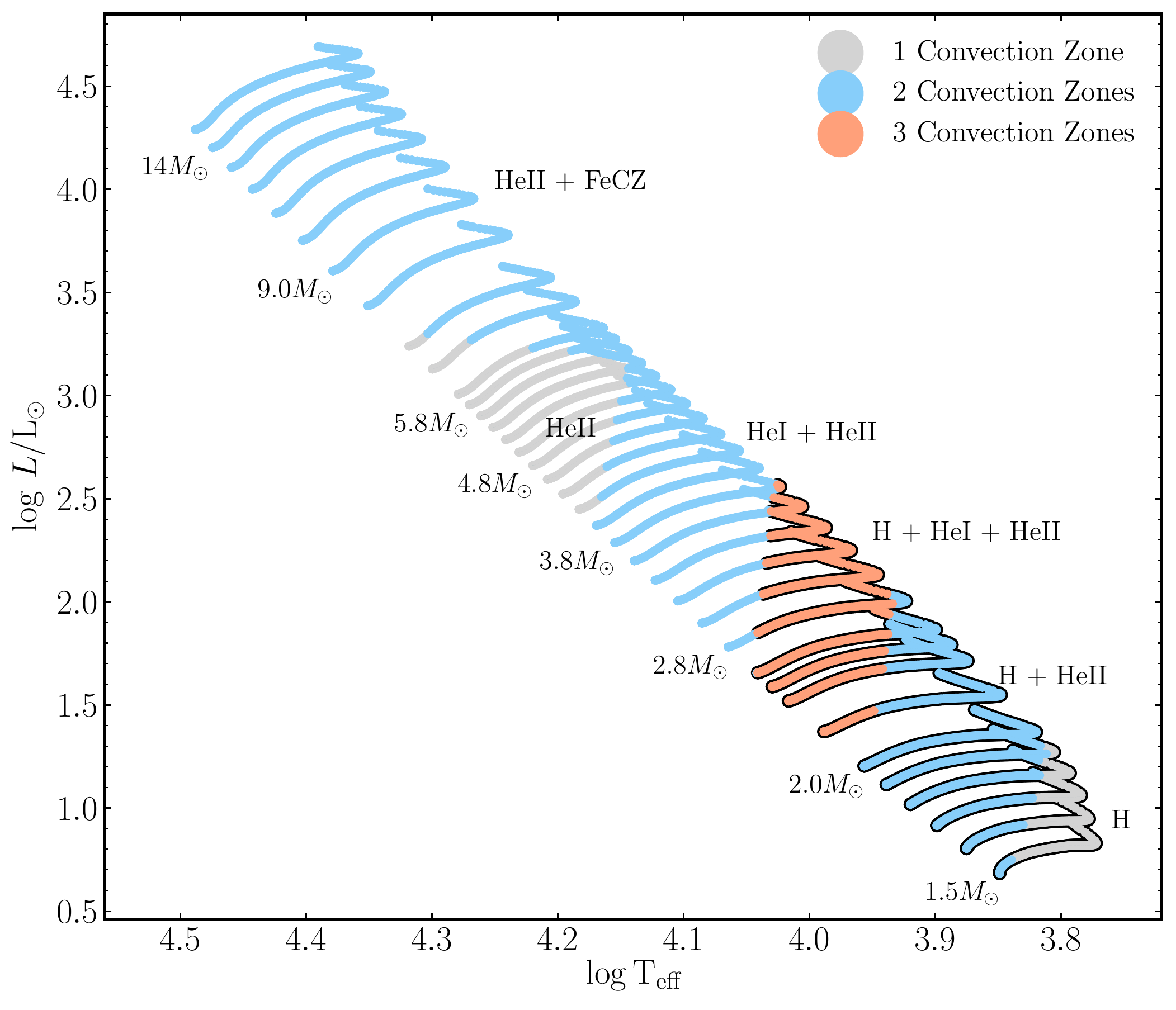}
\caption{{Number of main sequence envelope convection regions in stars between 1.5 and 15$\mso$ on the \hrd. Different colors show the number of co-existing convective regions present in the stellar outer layers. The type of convection zones is annotated on the plot. The black outline shows models where convection is present at the stellar surface. \label{HRdiagram}  
}}
\end{center}
\end{figure}

\begin{table*}
\begin{minipage}{\textwidth}
\centering
\caption{{Properties of the outer layers in non-rotating 2$\mso$, 2.4$\mso$ and 4$\mso$  main-sequence models with solar metallicity. 
}}
\label{table}

\begin{tabular}{ c c c c c c c c  c c c c c c}
\hline\hline
   $M_{\rm{ini}}$    & $R_{\star}$  & X$_{c}$\footnote{Core hydrogen fraction.}  \T  & $R_{\mathrm{HeII}}$\footnote{Radial coordinate of the top of the HeIICZ.} & $\Delta R_{\mathrm{HeII}}$\footnote{Radial extension of the HeIICZ.}    &      $\bar{\hp}$\footnote{Pressure scale height at top/bottom of the HeIICZ.} & $\vca$\footnote{Maximum of the convective velocity inside the HeIICZ.}  & $\bar{\vca}$\footnote{Average convective velocity inside the HeIICZ.}  & $\log \bar{\rho}$\footnote{Average density in the HeIICZ.} & $\log M_{\rm HeII}$\footnote{Mass contained in the convective region.}  & $\log M_{\rm top}$\footnote{Mass in the radiative layer between the stellar surface and the upper boundary of the convective zone.}    & $\tau_{\rm turn}$\footnote{Convective turnover time, $\tau_{\rm turn}:= 2\hp/<\vca>$.} &   $\tau_{\rm top}$\footnote{Time it takes to remove the material in the radiative layer, $\tau_{\rm top}:= \Delta M_{\rm top}/\dot M$.} & $\log \dot{M}$ \\
 $\mso$ &  $\rso$ &   & $\rso$   & $\rso$ & $\rso$ &  km~s$^{-1}$ &  km~s$^{-1} $     & g~cm$^{-3}$         & $\mso$             & $\mso$                            &       hours          &  years     & $\mso \rm{ yr}^{-1}$ \\

\hline
2.0 & 1.91 & 0.5 & 1.886 & 0.007 & 0.006 & 1.76 & 0.75 & -7.13 & -8.38 & -8.766 & 2.89 & 312 & -11.26 \\
2.4 & 2.36 & 0.47 & 2.330 & 0.009 & 0.007 & 1.07 & 0.42 & -7.29 & -8.263 & -8.653 & 6.41 & 401 & -11.26 \\
4.0 & 3.04 & 0.5 &3.020 & 0.009 & 0.007 & 0.16 & 0.06 & -7.68 & -8.401 & -8.826 & 48.38 & 21 & -10.14 \\
\hline

\end{tabular}
\end{minipage}
\end{table*}

\subsection{Envelope Convection}\label{subsurface}
At or near the surface of the star where abundant elements have  ionization thresholds, bumps in the heat capacity and opacity do indeed result in convective layers.
We call these HCZ, HeICZ, HeIICZ and FeCZ according to the species ionized there. 
We use the Modules for Experiments in Stellar Evolution (MESA, release 11342) to evolve non-rotating stellar models with masses between 0.9 and 25$\mso$ from the pre-main-sequence to close to the end of core H-burning\footnote{Defined as the point when the center mass fraction of H  is X$_{c} = 10^{-3}$} \citep{Paxton:2011,Paxton:2013,Paxton:2015,Paxton:2018,Paxton:2019}. The models have an initial metallicity of $Z=0.02$ with a mixture taken from \citet{Asplund:2005}.  Convective regions are calculated using the mixing length theory (MLT) in the \citet{Henyey:1965} formulation with $\alphamlt=1.6$, though we do check in section \ref{alphadependency} how our results depend on the $\alphamlt$ parameter. The boundaries of convective regions are determined using the Schwarzschild criterion.  An exponentially decaying overshoot above and below convective regions is accounted for with a parameter $\alphaov=0.014$ \citep{Herwig:2000,Paxton:2011}. Since we are just focusing on the convective properties of A and late B-type stars, we do not include the effect of stellar winds, which are basically negligible for the main-sequence evolution of these stars

The location of convection in solar-metallicity main-sequence stars between 0.9 and 25 $\mso$ is illustrated in Fig.~\ref{Kipp}. 
In stars above about 1.5 $\mso$ there is no thick convective envelope, just various thin convective layers, each occurring at a particular temperature. At low temperatures, ionization of H and HeI together produce one surface convective layer. Moving from lower to higher masses, each convective layer disappears as the photospheric temperature rises above each ionization temperature. This manifests itself in the locations on the \hrd of the boundaries separating stars with different numbers of convective layers, as seen in Fig.~\ref{HRdiagram}.
In our 1D models surface convection completely disappears around a temperature 10-11,000 K, with hotter models showing only subsurface convection zones.

Solar-metallicity stars more massive than about 7~$\mso$ have an iron-ionization convective layer, the FeCZ \citep{2009A&A...499..279C}. 
In contrast to the other envelope convection zones (HCZ, HeICZ and HeIICZ) which are always present when their relevant ionization temperatures occur inside the star, the FeCZ occurs only in more massive stars, even though the ionisation transition zone at 1.5$\times 10^5$ K is present in all stars. This can be explained as follows. The LHS of equation (\ref{eqn:schwarzschild2}) is essentially the same function of temperature in all stars since opacity and therefore also $\nabla_\kappa$ depend almost exclusively on temperature and only very weakly on pressure, and because (as discussed above) $\nabla_F-\nabla_g\approx0$. The difference between intermediate-mass stars and massive stars is on the RHS -- in more massive stars the radiation pressure $P_{\rm rad}=(a/3)T^4$ accounts for a greater fraction of the total pressure $P$, the specific heat capacities are correspondingly greater. To be more precise, in the absence of ionisation effects the value of $\nabla_{\rm ad}$ goes from $2/5$ in the limit of a monatomic ideal gas towards $1/4$ as radiation pressure becomes dominant. And from equation (\ref{eqn:schwarzschild1}) we see that the quantity $gT^4/FP$ is also the same function of temperature in all stars, which can be rearranged to give the radiative pressure fraction as $P_{\rm rad}/P \propto T_{\rm eff}^4/g_{\rm surf} \propto L/M$, where the latter is a measure of proximity to the Eddington limit. Plotting stellar models of fixed composition on an \hrd, the threshold for the existence of the FeCZ is indeed a line of constant $T_{\rm eff}^4/g_{\rm surf}$, which is approximately horizontal -- convection is present for a luminosity above about $10^{3.2}\lso$ at $Z=0.02$ \citep[Fig.~11 in][]{2009A&A...499..279C}.

Note that all of the subsurface convective layers are around 1 to 1.5 pressure scale heights in thickness. The reason for this can be seen in Fig.~\ref{Convection}: the opacity peaks are around 0.15~dex wide in temperature, and in pressure they have a width a factor $1/\nabla$ greater than this. Given that $\nabla\approx2/5$, this corresponds to a factor of $e$ in pressure or a little more.

\begin{figure}[h!]
\begin{center}
\includegraphics[width=1.00\columnwidth]{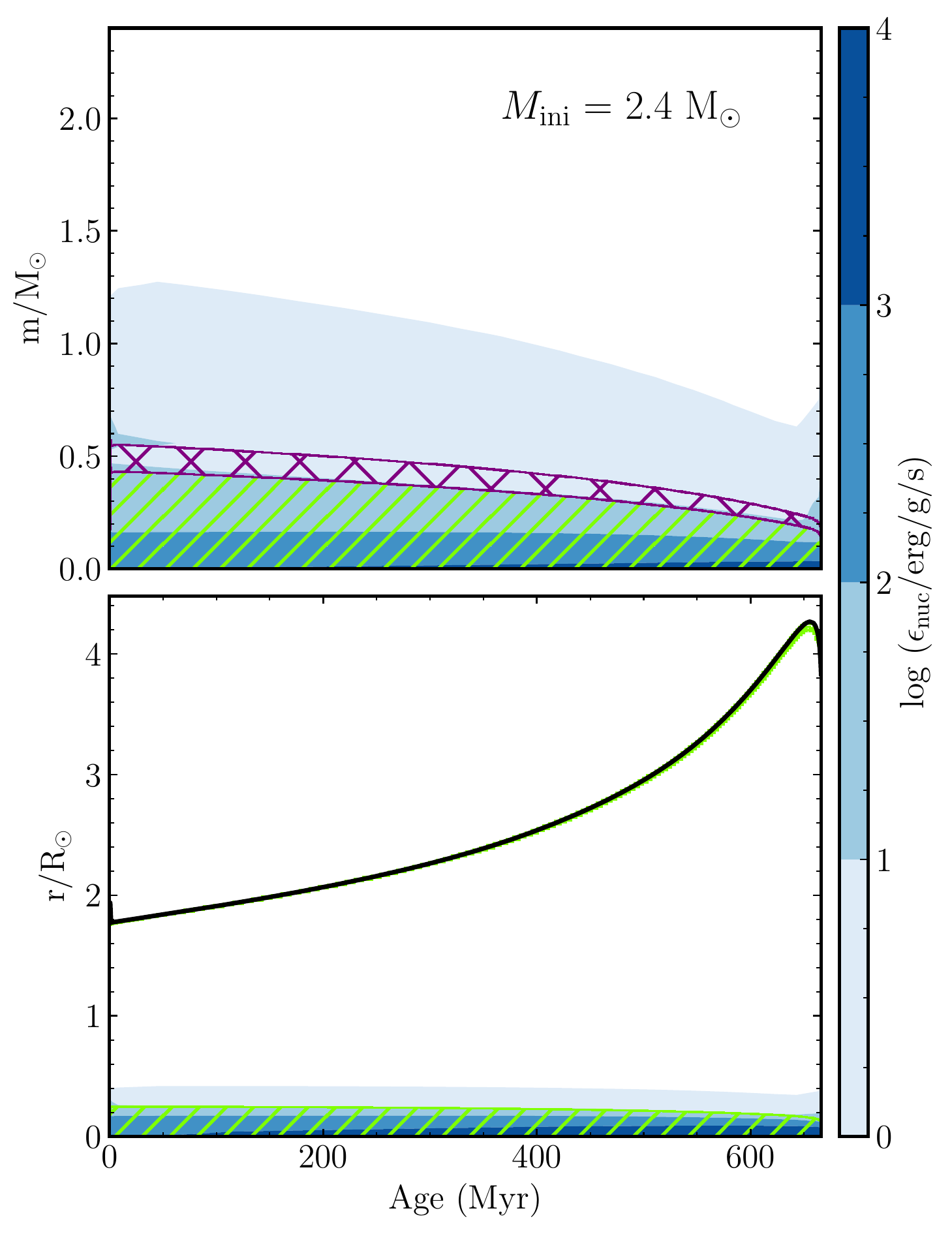}
\caption{{Top: Evolution of the internal  structure of a 2.4$\mso$ non-rotating star model during the main sequence. The Lagrangian mass coordinate is shown as function of stellar age. Convective regions are hatched green, overshoot regions are crossed purple, and regions of nuclear energy generation are shown as blue shading. Only the convective core is visible in this ``classic'' Kippenhahan plot. Bottom: same as above, but showing the evolution of the radial coordinate. The envelope convective regions are barely visible very close to the stellar surface. A zoom in is able to reveal their structure in Fig.~\ref{kipp_radius_n_vega}. \label{kipp_vega}
}}
\end{center}
\end{figure}

\begin{figure}[h!]
\begin{center}
\includegraphics[width=1.00\columnwidth]{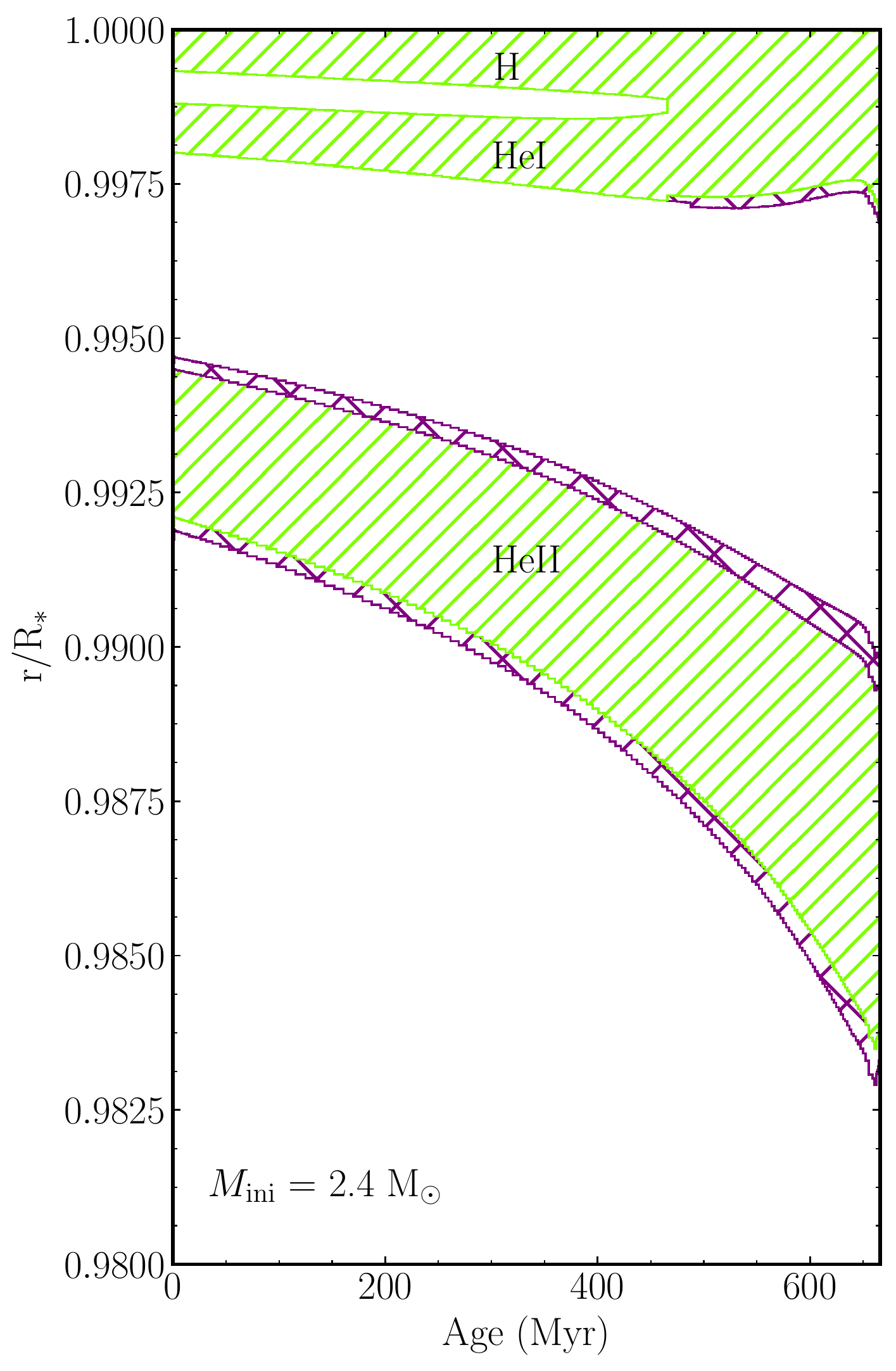}
\caption{{Evolution of envelope convection zones in a 2.4$\mso$ star model during the main sequence. The normalized radial coordinate is shown as function of stellar age. We plot the outer 2\% of the star to highlight the small envelope convective regions that are driven by partial ionization of H, HeI and HeII, respectively. Green hatched regions are convective, purple crossed regions show exponentially-decaying overshooting. In the calculations, the H and HeI convection regions merge at $t \simeq 460$ Myr. \label{kipp_radius_n_vega}
}}
\end{center}
\end{figure}

As stars move along the main sequence, convective layers can appear, disappear, merge and demerge. In Fig.~\ref{kipp_vega} and \ref{kipp_radius_n_vega} we illustrate the evolution of these layers in a stellar model with 2.4~$\mso$. Note that due to the very low densities in the outer layers ($\rho < 10^{-7}$ g cm$^{-3}$), these convective regions contain very little mass. As such, they are completely invisible in the usual Kippenhahan plot showing the evolution  of the internal structure in Lagrangian mass coordinate (See e.g. Fig.~\ref{kipp_vega}).  

In terms of the convective kinetic energy density, the most energetic of the layers is the deepest one, which between about 1.5 and 7 $\mso$ is the HeIICZ. 
In Table~\ref{table} we show the properties of the HeIICZ and the overlying radiative layer (ignoring the presence of the weaker convective layers) in 2, 2.4 and 4 $\mso$ models during core H-burning.

\begin{figure}[ht!]
\begin{center}
\includegraphics[width=1.00\columnwidth]{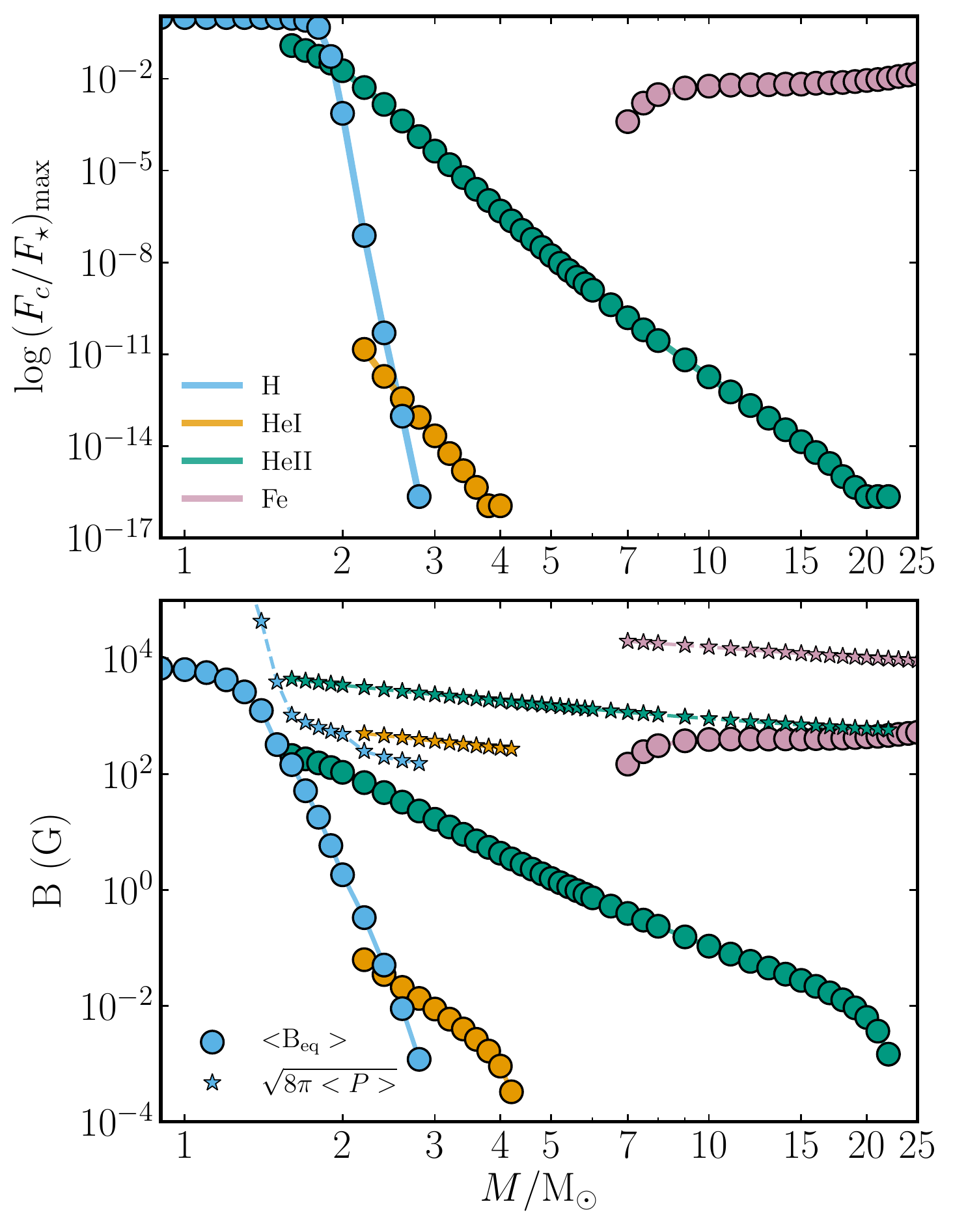}
\caption{{Top: Values of maximum convective flux in different convection zones during the main sequence (X$_c$=0.5) The flux is normalized to the total stellar flux. The transition from strong to weak convection occurs around 1.5-2.0$\mso$. Bottom: Values of the average equipartition (with kinetic energy) magnetic fields in the various (sub)surface convection zones. For reference we also show values of the equipartition  with total pressure magnetic field. The adopted value of total pressure used is an average calculated across the convective zone. \label{convective_efficiency}
}}
\end{center}
\end{figure}

\subsection{Weak convection}
In the HeIICZ the transport of energy by convective motions is relatively inefficient: radiation dominates and transports more than 95\% of the total flux. This is because the density is very low and the mean free path of photons correspondingly long. In this situation convection is significantly superadiabatic, and the gradient $\grad\equiv {\rm d}\ln T/{\rm d}\ln P$ is explicitly calculated from the mixing length equations \citep[e.g.][]{Kw90}.

The convection in the HeICZ and photospheric HCZ is even weaker than in the HeIICZ. The calculations presented so far have ignored the role of viscosity, as is common practice in stellar modelling, using simply the Schwarzschild criterion appropriate for convection at high Reynolds number. In a viscous fluid the instability condition can be expressed in terms of the so-called Rayleigh number Ra as
\begin{equation}
{\rm Ra}\approx\frac{(\nabla-\nabla_{\rm ad})L^3g}{\chi\nu}>{\rm Ra}_{\rm crit}\approx10^3 \label{Eq:Ra}
\end{equation} 
where again $\nabla$ and $\nabla_{\rm ad}$ are the actual and adiabatic temperature gradients,
$\chi$ an $\nu$ are the thermal and kinetic diffusivities, $g$ is gravity and $L$ is some relevant length scale in the vertical direction, presumably either the vertical extent of the convective zone or the scale height if that is smaller. 
The value of ${\rm Ra}_{\rm crit}$ can be determined experimentally but varies by a factor of 2 or 3 depending on the boundary conditions. Note that the presence of rotation also alters the value of ${\rm Ra}_{\rm crit}$.  We calculated the Rayleigh number in each convective layer in a $2.4\mso$ model with an effective temperature T$_{\rm eff}=9560$ K (we chose this specific model to reproduce the conditions of Vega, a well-known A-type star). Using the numbers at the location in each layer of peak convective energy density $\rho \vca^2/2$, we find Ra $\approx 10^2$, $10^3$ and $10^7$ in the HCZ, HeICZ and HeIICZ respectively. 
The reason for the difference in Ra between the layers is largely a consequence of the density: at lower density the mean free path of the photons, which carry both heat and momentum, is greater and both thermal and kinetic diffusivity are higher. Indeed at the photosphere the mean free path is comparable to the scale height. In addition, the scale height used in the numerator in Eq.(\ref{Eq:Ra}) is smaller closer to the surface.
 
It is not well understood how  convection should look at low Rayleigh number; there should perhaps be some kind of viscous, non-turbulent motion.
Another uncertainty is our limited inability to predict the temperature gradient and convective velocities accurately; we currently use mixing length theory. Although standard in stellar evolution modelling, we see from laboratory experiments and numerical simulations, as well as the experience of glider pilots, that mixing length theory is based on an inaccurate physical picture. In reality rising and falling fluid parcels move past each other rather than mixing, surviving over many scale heights; heating and cooling at the boundaries is crucial. In the absence of a quantitative theory based on real physics we use the MLT in the modelling; the numbers should therefore be treated as the result of a dimensional analysis, and their dependence on the mixing length free parameter $\alphamlt$ should neither worry nor surprise us. In the literature a number of works have attempted at simulating these convective regions using multi-dimensional hydro simulations, although mostly for cool A stars with surface temperatures below 8,500 K \citep[see e.g.][]{Kupka:2002,Trampedach:2004,Kochukhov:2007,Freytag:2012,Kupka:2017}. For the more inefficient convective regions arising in hotter A and early B-type stars, the situation still needs to be clarified. This could be done with the aid of state-of-the-art radiation hydro simulations, similar to those of \citet{Jiang2015} but for much lower values of the stellar luminosity.

\begin{figure}[ht!]
\begin{center}
\includegraphics[width=1.00\columnwidth]{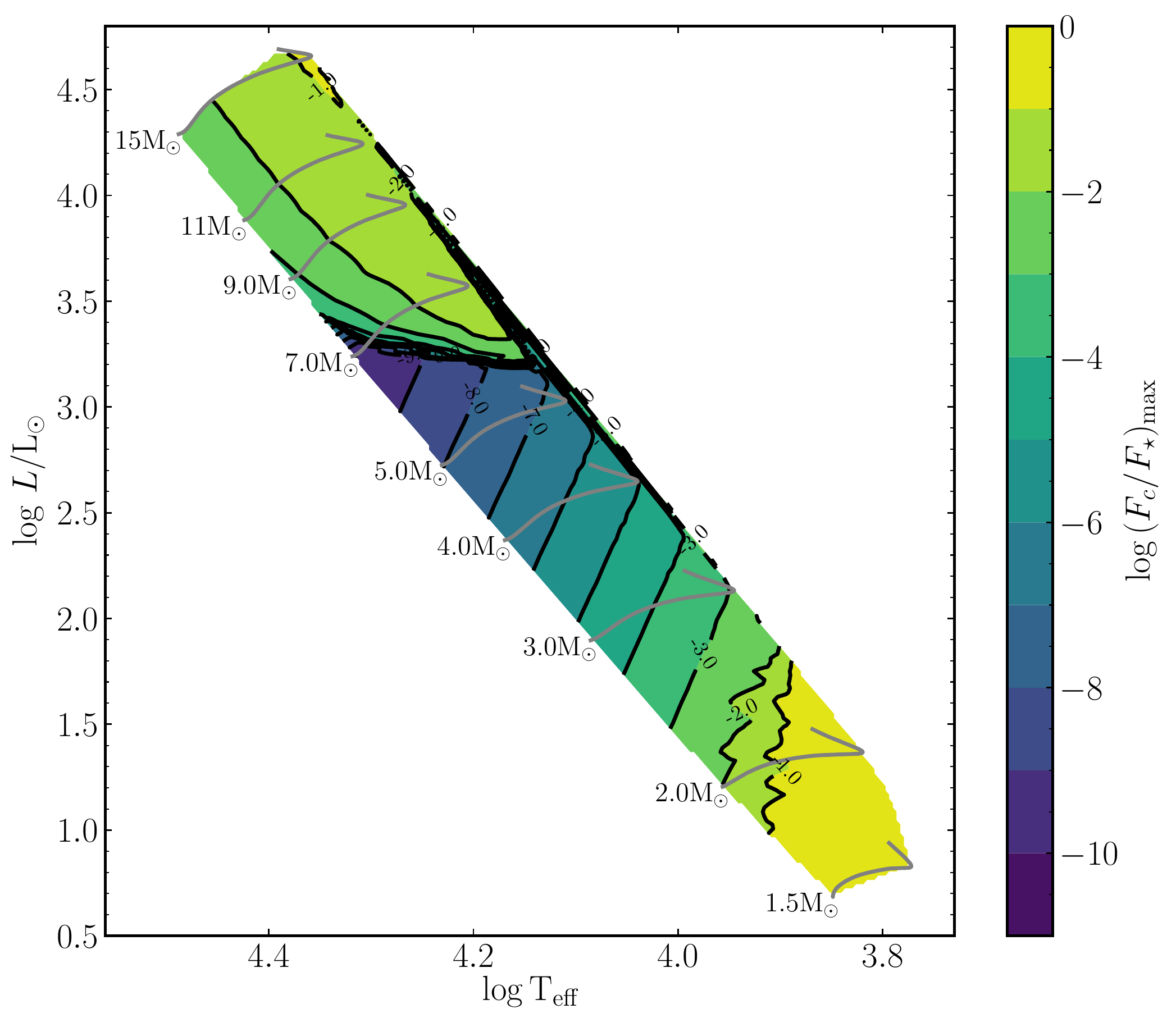}
\caption{Values of the maximum convective flux in any convection zones during the main sequence. The flux is normalized to the total stellar flux. In the lower right corner the HCZ is the most efficient convective zone, while at higher luminosity the FeCZ is the  subsurface convective region transporting the largest fraction of the stellar flux. In the rest of the \hrd, the HeIICZ is the most efficient envelope convection region, despite transporting a very small fraction of the stellar flux. \label{convective_efficiency_hrd} }
\end{center}
\end{figure}

 Observationally, the transition from convective to radiative surfaces seems to occur at surface temperatures between 9,000 and 10,000 K, depending on the proxy. This is marginally consistent with the results shown in Fig.~\ref{Kipp} and Fig.~\ref{HRdiagram}, where surface convection as calculated using the MLT disappears at around 10,000-11,000 K.
 \citet{2009A&A...503..973L} find microturbulent velocities in the photospheres of stars to be approximately 2~$\kms$ in stars cooler than 10,000 K, while above that temperature there is an upper limit of around 1 $\kms$. The microturbulent velocity is interpreted as convective line broadening, but also internal gravity waves coming from the HeIICZ could contribute \citep{2009A&A...499..279C}.
 Chromospheric activity indicators and large line profile  asymmetries are also present in stars cooler than about 8,250 K ($\log $T$_{\rm eff} < 3.92$ ) \citep{Simon:2002,2009A&A...503..973L}, a temperature that coincides pretty well with the transition to largely inefficient surface H convection (see Fig.~\ref{convective_efficiency} and \ref{convective_efficiency_hrd}).

\section{Magnetism}\label{spots}
The presence of convection zones close to the surface of A~stars opens the possibility that magnetic fields could be produced by dynamo action and reach the surface, for example via magnetic buoyancy. Below we discuss this hypothesis.
\subsection{Dynamo Action}\label{dynamo}
 In an astrophysical plasma a dynamo is a configuration of the flow which is able to generate a magnetic field and sustain it against Ohmic dissipation. Depending on the scale of the resulting magnetic field with respect to the scale of kinetic energy injection, dynamos can be divided into small and large scale. In large scale dynamos the field has correlation length bigger than the forcing scale in the flow, while small scale dynamos result in magnetic fields with correlation scale of order of or smaller than the forcing scale. 
 It is generally considered that large-scale dynamos require fast rotation, characterised by a low Rossby number (the ratio of rotation period to convective turnover time), as well as perhaps differential rotation.

In principle the HeIICZ could host a small-scale dynamo. Moreover, most intermediate-mass stars rotate rapidly, with rotation periods of 12 hours or 1 day, which corresponds to a rotational period of the order of the convective turnover timescale (Rossby number is in the range 1...10), so it may be possible to build magnetic structure on a larger length scale than that of the convective motion.
Assuming the dynamo produces a magnetic energy density equal to the convective kinetic energy density, we calculate magnetic field strengths up to a few hundred gauss inside the HeIICZ -- see Fig.~\ref{convective_efficiency}. 

This assumption is supported by numerical simulations showing dynamo excitation by convection in the presence of rotation and shear, which also show magnetic fields reaching equipartition on scales larger than the scale of convection
\citep{2008A&A...491..353K,2010arXiv1009.4462C}. 

\begin{figure}[h!]
\begin{center}
\includegraphics[width=1.0\columnwidth]{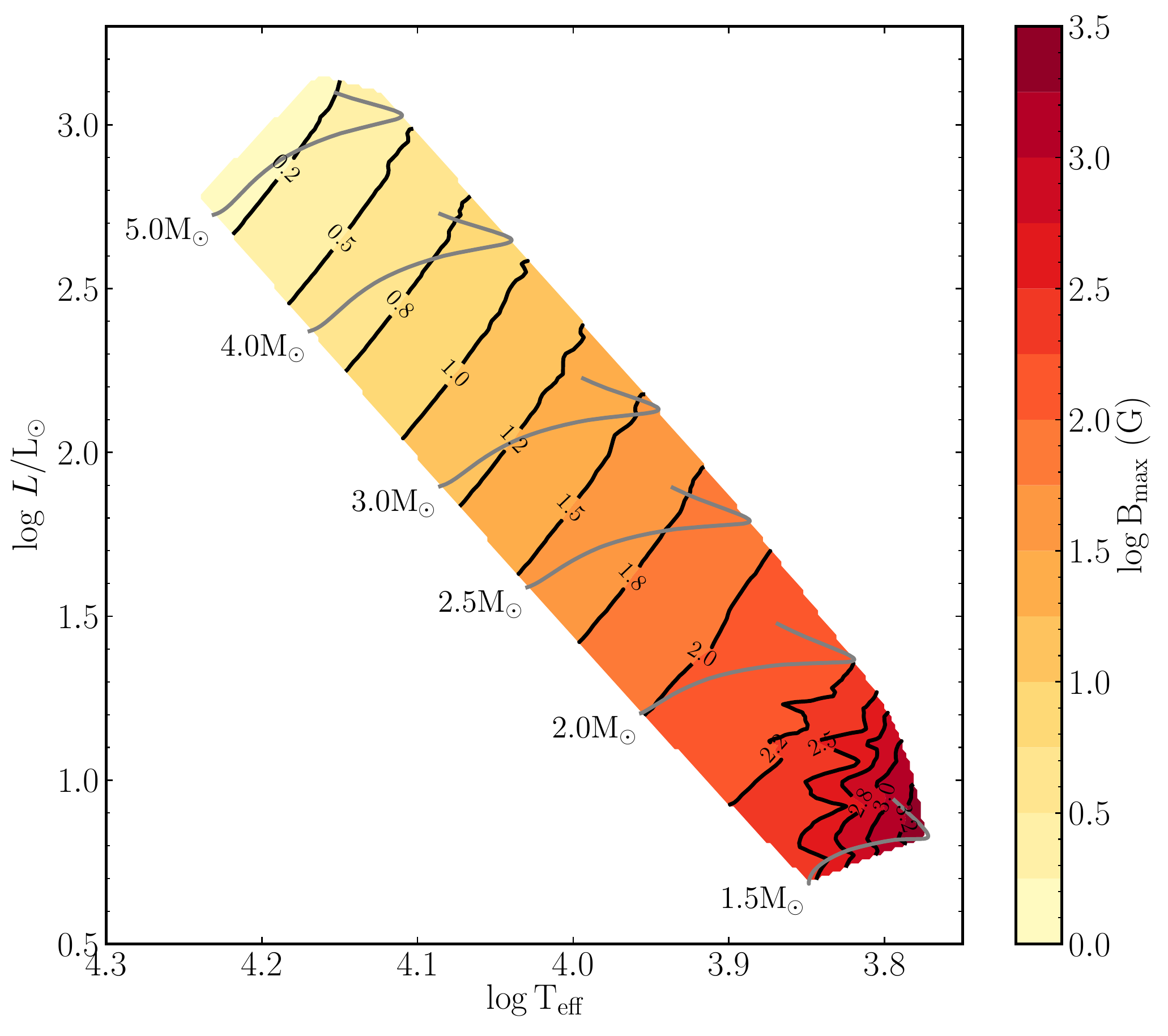}
\caption{Maximum magnetic field as function of location in the \hrd in any envelope convection zone calculated assuming equipartition between convective energy density and magnetic energy density. This plot is based on main-sequence evolutionary models between 1.5$\mso$ and 5$\mso$ with Galactic metallicity (Z=0.02). The convective properties are calculated using the mixing length theory. Some evolutionary tracks are shown for references. \label{beq_AB}}
\end{center}
\end{figure}

Hence, the nature of the dynamo action in the HeIICZ  could depend on parameters like the stellar rotation and the shear profile. The dynamo may also be affected by a fossil or failed fossil large-scale magnetic field penetrating upwards from the radiative zone below (see Sect.~\ref{disc}). In Fig.~\ref{beq_AB} we show the expected maximum magnetic field inside any envelope convection region, assuming equipartition of magnetic and kinetic energy:
\begin{equation}
\label{eqn:equipartition}
\frac{B^2_{\rm eq}}{8\pi} = \frac{1}{2}\,\rho\,\vca^2,
\end{equation}
where we adopted the maximum value of the $\rho \vca^2$ 
inside the convective zones as computed in the non-rotating models. Due to the higher power dependency on the velocity, the location of the maximum of $\rho \vca^2$ always roughly corresponds to the maximum of $\vca$. Values of magnetic fields calculated using the average convective velocity and density typically differ by less than $30\%$. For models in the range $\simeq$ 1.5 to 5$\mso$ and for  temperatures above $\teff \simeq 3.85$, the HeIICZ is the convective zone hosting the strongest equipartition magnetic fields. The HCZ contributes with the strongest fields, but only for the cooler models in the lower corner of the plot.

\subsection{Magnetic Fields Rise}\label{buoyancy} 

Now that we have established a substantial magnetic field can in principle be generated by dynamo action in the HeIICZ, we would like to know how it could rise through the overlying radiative layer and reach the photosphere. In \citealt{2011A&A...534A.140C} we compared various mechanisms to bring magnetic field upwards. Upwards advection by mass loss and Ohmic diffusion are both slow, and convective overshoot is unlikely to bridge the gap between the HeIICZ and the stellar photosphere (see e.g. Fig.~\ref{kipp_radius_n_vega}), but buoyancy can bring field to the surface on the dynamic (Alfv\'en) timescale.

A magnetic field provides pressure without contributing to density, so a magnetic feature which is in pressure equilibrium with its surroundings must have a lower temperature in order to have the same density and avoid rising buoyantly; the result is inwards diffusion of heat. In this context the mean free path of photons is large and heat diffusion keeps magnetic features at roughly the same temperature as their surroundings. Consequently it rises at a speed limited by aerodynamic drag, which works out to be 

\begin{equation}\label{eqn:drag}
v_{\rm drag} \sim  v_{\rm A} \left(\frac{l}{\hp}\right)^{1/2} \sim \frac{c_{\rm s}}{\beta^{1/2}}\left(\frac{l}{\hp}\right)^{1/2} 
\end{equation}
where $l$ is the size of the magnetic feature and $c_{\rm s}$ is the sound speed. Note that this buoyant rise happens at essentially the same speed as the adiabatic magnetic buoyancy instability \citep{1961PhFl....4..391N,1966ApJ...145..811P}, which may also be relevant here. For a magnetic feature with $l\sim \hp$ and $\beta\approx500$, the time taken to rise one scale height is of order 5 hours in our Vega model.

The difference between the field strengths at the top and bottom of the radiative layer depends on its geometry. In a self-contained magnetic feature (a `blob' or `plasmoid') the field strength scales as $B\propto \rho^{2/3}$ as it rises. Given that $P\propto\rho^{4/3}$ in the radiative layer (approximately, since $\nabla\approx1/4$), we see that $\beta$ remains constant during the rise. Such plasmoids might however be difficult to detach from the convective layer. An alternative geometry would be a horizontal flux tube, where the scaling is $B\propto\rho$. More realistic is a sunspot-like arch, where the central section of the tube rises to the surface while its ends are still in the convective layer, allowing plasma to flow downwards along the tube. Since the temperature inside and outside the tube are essentially the same, the pressure scale heights are also roughly equal inside and outside.

\begin{figure}[h!]
\begin{center}
\includegraphics[width=1.0\columnwidth]{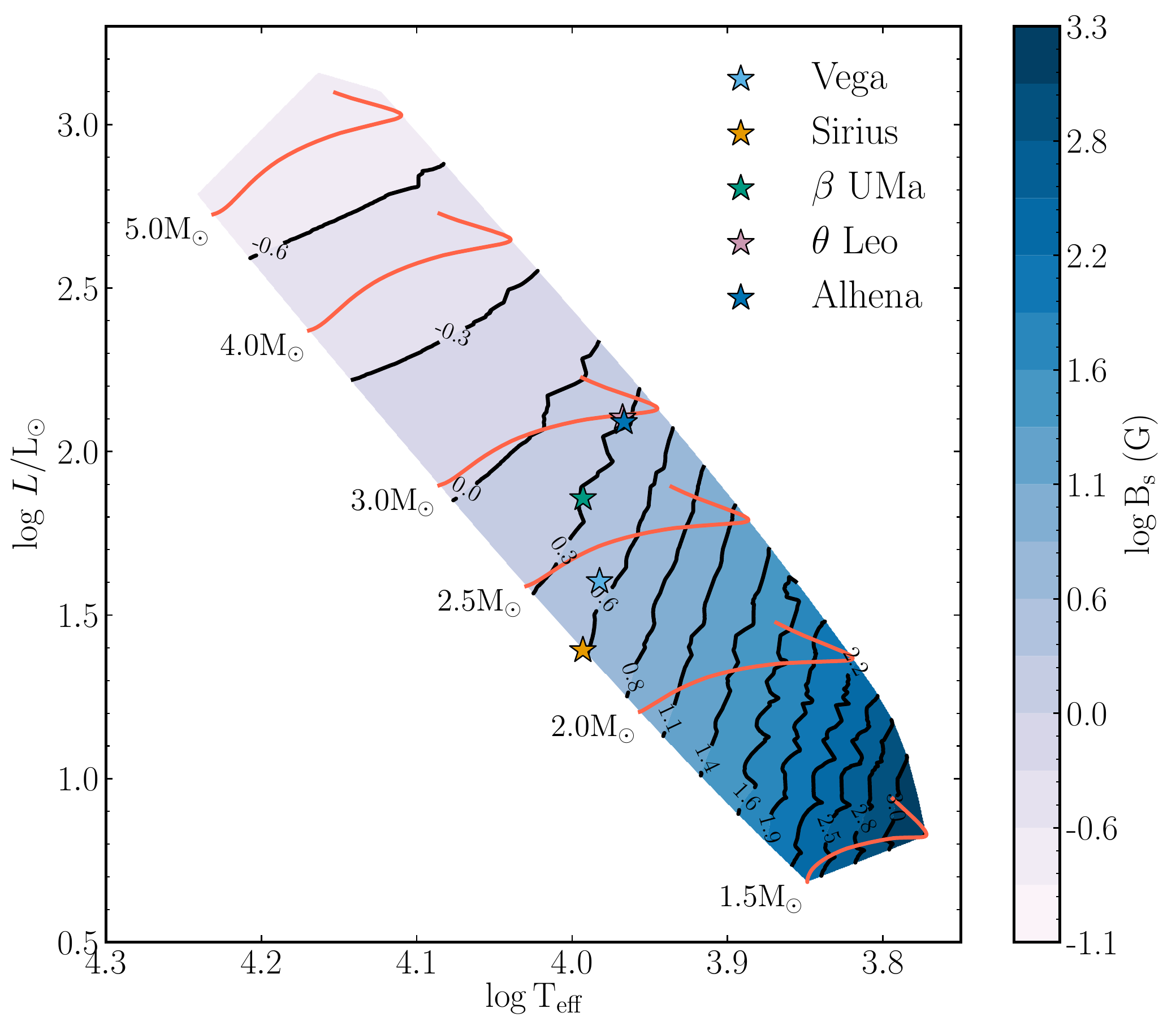}
\caption{Values of expected surface magnetic fields (in gauss) as a function of location in the \hrd. This plot is based on main-sequence
evolutionary models between 1.5$\mso$ and 5$\mso$ with Galactic metallicity. Some evolutionary tracks are shown for references. The surface magnetic fields are calculated scaling the equipartition fields in the envelope convection zones (see Fig.~\ref{beq_AB}) as $\rho^{2/3}$ (see discussion in Sect.~\ref{consequences}). The star symbols correspond to the locations of known magnetic (non Ap) A-stars, having surface magnetic fields of the order $\sim1-5$~G. 
For these stars the predicted surface magnetic fields from subsurface convection is $\sim2-4$~G. The stars $\theta$~Leo and Alhena have almost identical positions in the \hrd. \label{bsur_AB}}
\end{center}
\end{figure}

This means that the plasma-$\beta$ is equal along the length of the tube: we have the same scaling with density $B\propto \rho^{2/3}$ as with the plasmoid scenario.
In Fig.~\ref{bsur_AB} we show the expected amplitude of surface magnetic fields, calculated assuming buoyant rise of the equipartition fields shown in Fig.~\ref{beq_AB} (a scaling $B\propto \rho^{2/3}$ as the feature rises).

Note that there could be other ways for a magnetic field to escape the subsurface convection region and reach the stellar surface. For example, \citet{2010A&A...523A..19W,2011arXiv1104.0664W} studied the magnetic flux produced by a turbulent dynamo in Cartesian and spherical geometry, respectively. They found that magnetic flux can rise above the turbulent region without the need of magnetic buoyancy. This appears to be related to the release of magnetic tension, which leads to the relaxation and emergence of the field. In this case the magnetic field at the surface could be larger than in the case of magnetic buoyancy. Therefore, the values of surface magnetic fields reported in Fig.~\ref{bsur_AB} should be taken as lower estimates (but see also the discussion on the dependency of the equipartition magnetic fields on $\alphamlt$ in Sec.~\ref{alphadependency}). 
Even in the absence of a dynamic process bringing the dynamo-generated magnetic fields to the surface, it is important to notice that the radiative layer separating the HeIICZ from the surface contains very little mass. Even the very weak stellar winds of A-type stars can easily remove this mass on a timescale of a few hundred years or so (see Tab.~\ref{table}). Therefore the material present at the surface of an A star has been recently stirred by turbulent convection. Magnetic fields produced by a convective dynamo  tend to decay rapidly (on an Alfv\'en timescale) when they are not constantly regenerated, but in the presence of the rapid rotation typical of A stars, this process can be delayed substantially by the stabilizing effect of the Coriolis force \citep{2013MNRAS.428.2789B}. Hence, the stellar plasma might still be substantially magnetized by the time the layers are revealed to the surface by  stellar winds.

\subsection{Magnetic Field Variability}\label{variability}
Once a magnetic feature reaches the surface of the star 
it would be useful to estimate its lifetime $\tau_{\rm spot}$. 
As a lower limit for $\tau_{\mathrm{spot}}$ we could take the Alfv\'en crossing time, or equivalently the time a feature of size $l$ takes to cross the photosphere while rising with a velocity $v_{\rm drag}$. For $l \sim \hp$ this gives a timescale of the order of a few hours. 
Unlike in more massive stars, the upper limit to the lifetime from the removal of the spot by the loss into the wind of the gas it is threading is very long, since the mass-loss rate is very small. A more relevant upper limit is probably set by the dynamo properties.  Magnetic features should probably persist for at least the smallest imaginable dynamo timescale, the convective turnover time, which is less than a day. However, with the aid of rotational effects, greater dynamo timescales could be present. If large scale magnetic fields are produced, these could be stable on much longer timescales, akin to the long timescales of field-reversals observed in some stars ($\sim$ years). A co-existence of short-lived and long-lived magnetic features is also possible, and required to explain recent spots evolution observations of \citet{Petit:2017}.

\section{Observable effects}\label{consequences}
\subsection{Surface Magnetism}
As far as direct detection of the magnetic field from spectropolarimetric observations of the Zeeman effect is concerned, the most readily measured quantity is the mean longitudinal field, i.e. the disc-averaged line-of-sight component. The detectability depends on the field strength and geometry. Unfortunately it is not obvious what the geometry might be. On the entire surface of the Sun we see magnetic structure on the granulation scale: a small-scale local dynamo located near the surface. If the A-star dynamo looks like this, and the field everywhere is able to escape upwards, the entire photosphere will be covered in randomly-oriented magnetic field with a length scale comparable to the scale height in the HeII zone, $H_{\rm c}$. Assuming $\sqrt{N}$ statistics in the disc-averaging, the observed mean longitudinal field would be
\begin{equation}
B_{\rm long} \sim \frac{H_{\rm c}}{R_\ast}B_{\rm surf}
\end{equation}
where $B_{\rm surf}$ is the surface field strength as calculated in the previous section. In our Vega model $B_{\rm surf}\sim$5 gauss and $B_{\rm long}\sim B_{\rm surf}$/400, which is completely unobservable with current technology and also incompatible with the measured value longitudinal field of $0.6 \pm 0.3$ gauss \citep{2009A&A...500L..41L}. Actually even $\sqrt{N}$ statistics seems optimistic since it assumes the polarity of a region is random and independent of its neighbours, which is unlikely to be the case.

Perhaps though the length scale at the surface is larger than $H_{\rm c}$, since magnetic features expand as they rise towards the stellar surface; their size would be greater by a factor $(\rho_{\rm c}/\rho_{\rm ph})^{1/3}$. Unfortunately this is still insufficient to explain the observations.

It {\it is} possible however for dynamos to produce structure larger than the convective scale. In terms of observational evidence, we see that fully convective stars generally have strong dipolar fields, but that stars with a convective envelope such as the Sun have weaker dipole fields and that most of the energy is at smaller scales \citep[see e.g.][]{2010MNRAS.407.2269M,2012ApJ...755...97G}. Intuitively this is understandable in terms of the difficulty in different parts of a convective shell `communicating' with each other to produce coherent large-scale structures. Probably the thinner the shell is, the more serious is this problem. Differential rotation probably helps though to connect different parts of a convective shell, probably most effectively in the azimuthal direction. Sunspots are produced by some process around the base of the convective envelope; the active regions we see at the surface are no larger than the scale height at that depth. With only a very thin convective layer to work with, it is difficult to imagine producing large active regions or spots on an A star.

\subsection{Photometric Variability}
\begin{figure}[ht!]
\begin{center}
\includegraphics[width=1.0\columnwidth]{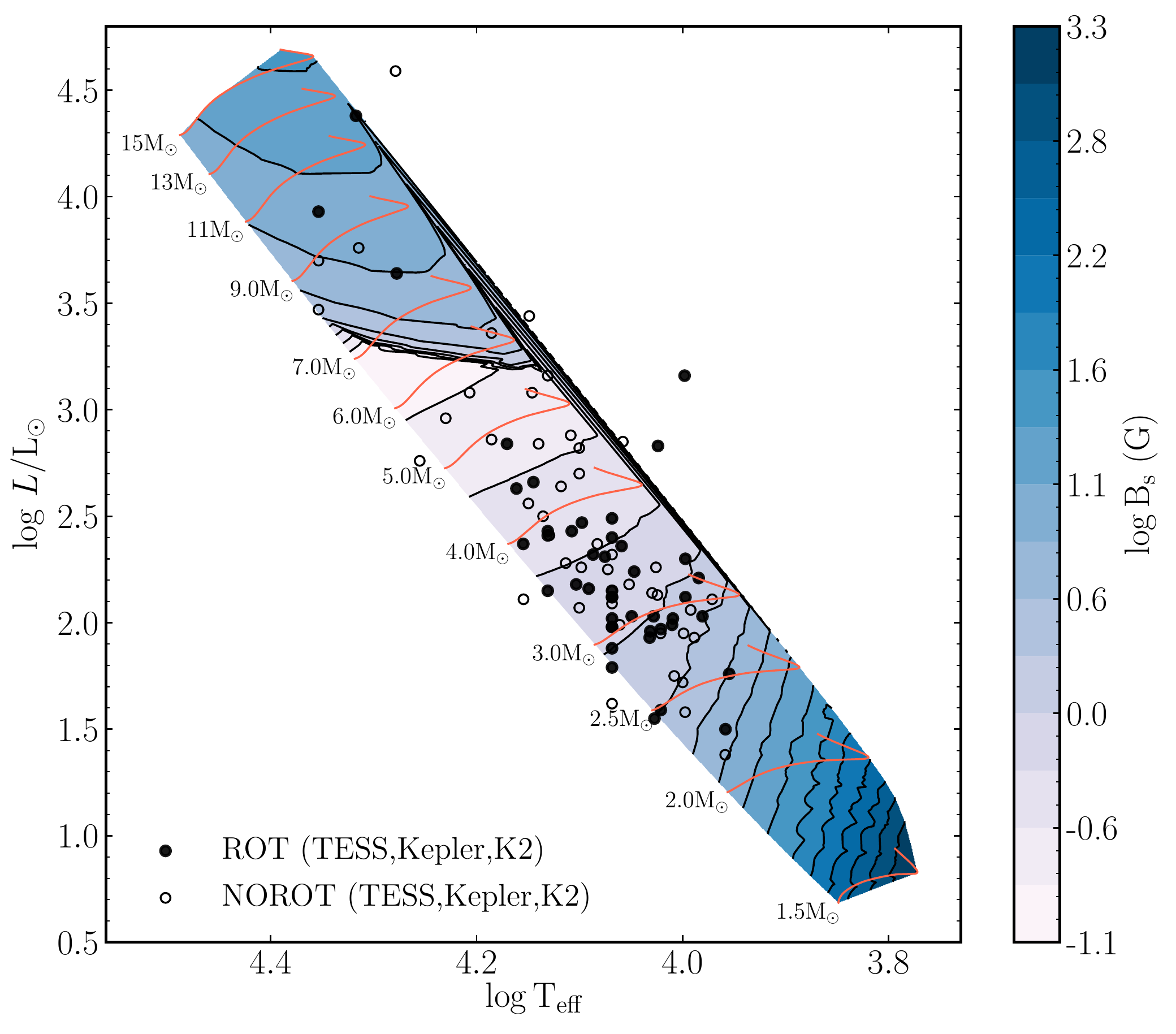}
\caption{Values of expected surface magnetic fields (in gauss) as a function of location in the \hrd. This plot is based on main-sequence evolutionary models between 1.5$\mso$ and 15$\mso$ with metallicity Z=0.02. Magnetic fields produced by the FeCZ dominate the high-mass regime, while at lower luminosities magnetic fields are emerging from the HeIICZ (except for models with $\teff  \lesssim 3.9$, where the HCZ dominates). Observations from \citet{Balona:2019} are visible as circle symbols, with filled circles showing the location of stars identified as possible rotational variables (ROT).  \label{bsur_OBA}}
\end{center}
\end{figure}

Apart from the Zeeman effect, a magnetic field has various potentially-observable indirect effects. For instance, in general a magnetic field affects the temperature of a radiative photosphere. The photosphere is at a constant optical depth, but lower inside magnetic features because of the contribution of magnetic pressure. Since temperature is a function of height, magnetic spots appear hotter and brighter than their surroundings\footnote{Note the contrast to sunspots, which appear dark because the magnetic field inhibits convection and consequently also heat transport.}. The temperature difference is given by \citep{2011A&A...534A.140C}
\begin{equation}
\frac{\Delta T}{T}\approx\frac{\adrad}{\beta} \, .
\end{equation}\label{temp}
which in Vega is only $\sim10$ K if $T=10^4$ K and $\beta=500$, corresponding to a difference in flux density of less than 1\%. If the spot is larger than the photospheric scale height, transport of heat between outside and inside the spot is less effective and temperature is no longer purely a function of height, so the photospheric temperature difference is likely smaller. Whilst the effect might just be observable on more massive stars, it might be too small to explain the observations of \citet{Bohm:2015} of Vega, who find photometric variability which they model as a flux density difference of $5\,10^{-4}$ in a large spot with a radius equal of $0.3\,R_\ast$. Perhaps the observations could also be reproduced with a large number of smaller and somewhat stronger spots \citep{2017MNRAS.472L..30P}. 
Rotational modulation of surface structures akin to stellar spots has been reported in a large number of B and A stars observed by Kepler and TESS \citep{Balona:2011,Balona:2019,Pedersen:2019}. In Fig.~\ref{bsur_OBA} we show the location of possible rotational variable stars in \citet{Balona:2019}, together with the predicted surface magnetic fields coming from subsurface convective regions.  While it is difficult to make firm conclusions, it is interesting to note  a dearth of rotational variables candidates in the region corresponding to minimum predicted surface magnetic fields.  In the near future, more observations coming from TESS should help to further explore this correlation.   

\subsection{Flares}
There is some evidence of flares in Am/A stars \citep{Balona:2013}. Although normally one can invoke an otherwise undetected low-mass companion to explain flares (and also X-rays) \citet[see e.g.][]{Pedersen:2017}, there are at least a few cases where the flare energy is probably too large for this to be plausible. Explaining these energetic flares with a subsurface magnetic dynamo and associated coronal activity is also not easy. 

In the Sun, flares are believed to be produced by magnetic reconnection in strongly magnetized regions (sunspots) having scales $L \sim 0.01 \rso$ and $B\sim kG$. In the case of very efficient energy conversion, flares can have energies as high as $E_{\rm flare} \approx B^2 L^3 \sim 10^{32}$ erg.
In the case of an A star like Vega, where the magnetic field is about two orders of magnitude smaller, one needs magnetic fields concentrations on scales larger than the stellar diameter to explain the observed maximum energy $E_{\rm flare} \sim 10^{36}$  erg \citep{Balona:2012,Balona:2013}, which is clearly not possible.  Moreover, far ultraviolet observations of A stars have shown that chromospheric activity seems to disappear for surface temperatures above 8,250 K \citep{Simon:2002}. 

This said, given the typical flares recurrence time $\tau$ of 10-100 days, one can still accommodate them via a build-up of magnetic energy within the HeIICZ or in the radiative layer above it. For a typical A star, the convective luminosity $L_{\rm c}$ (the rate at which energy passes through the convective motion, of order $\rho v_{\rm c}^3$ per unit area) is a fraction $10^{-2}-10^{-3}$ of the total stellar luminosity (see Fig.~\ref{convective_efficiency}). Again, assuming equipartition of kinetic energy density with magnetic energy density and the ability of magnetic fields to store a small fraction $f$ of the integrated convective luminosity during the recurrence time, one can power flares with energy as large as
\begin{equation}
E_{\rm flare} \approx f \, \tau  L_{\rm c},
\end{equation}
which for a typical A-star only requires $f\sim 10^{-3}$ or so to reach $10^{36}$ erg. Of course this would imply that the flares in A stars are produced in a very different way than in Sun-like stars, something that needs further observational and theoretical studies.

\section{Vega}\label{sec:vega}

The prototype A0 star Vega is a rapid rotator, with extensive monitoring in  spectropolarimetry, photometry and interferometry. Vega is the first A star for which a weak surface magnetic was reported \citep{2009A&A...500L..41L}. The magnetic field has a $0.6\pm 0.3$ G disk-averaged line-of-sight component, with peak values of 7 G \citep{Petit14,Petit:2017}. The magnetic topology was reconstructed using Zeeman-Doppler imaging, which shows a prominent polar magnetic region and a few other magnetic spots at lower latitude.

\begin{figure}[h!]
\begin{center}
\includegraphics[width=1.00\columnwidth]{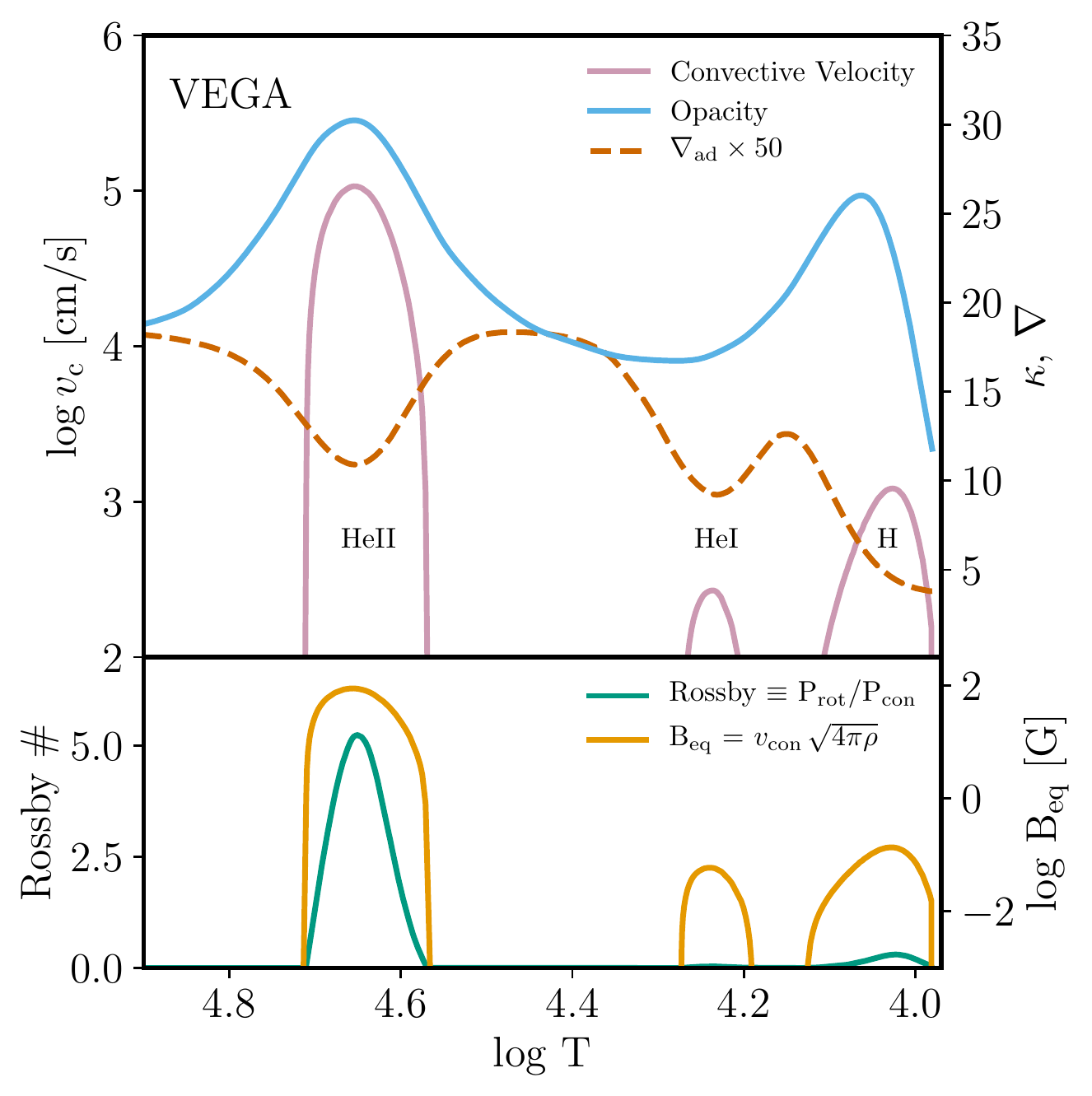}
\caption{{\label{Vegaconv} Top: Convective velocities as function of temperature in the outer layers of a $2.4\mso$ model with T$_{\rm eff}=9560$ K representative of the star Vega. The stellar surface is to the right. The velocities are in $\log$ (cm/s) and have been calculated using a value  of the mixing length parameter $\alpha_{\rm MLT} = 1.6$. We also show the opacity and label the convective zones according to their origin (partial ionization of He or H). Bottom: profiles of the Rossby number and the equipartition magnetic field in the outer layers of the model. The Rossby number is calculated assuming a rotation period of 13 hours, appropriate for the rapidly rotating star Vega.%
}}
\end{center}
\end{figure}

Looking at our Vega model in Tab.~\ref{table} ($M_{\rm{ini}} = 2.4\mso$), we can see that the turnover time in the HeIICZ is about 6 hours. Since the rotation period of Vega is only 0.68~d, the Rossby number is order unity (Fig.~\ref{Vegaconv}) and a convective dynamo in this envelope convection zone is expected to be efficient at creating an equipartition magnetic field. While this is possibly the case for the HCZ and HeICZ, in the context of the MLT these convective regions contain negligible kinetic energy densities and are of no interest for the problem of explaining the observed surface magnetic fields. At the location of Vega, the calculated model has $B_{eq}$=35...86 G in the HeIICZ, depending if we take average or max convective velocity. Assuming the usual $\rho^{2/3}$ \citep{2011A&A...534A.140C} scaling, buoyant surface magnetic fields can reach 2...5 G, pretty close to the maximum observed value of $\sim$ 7 G at the pole of the star \citep{Petit14}. It is not possible to say much about the geometry of the magnetic field, as this requires a careful study of dynamo action in the thin HeIICZ.

Strong evidence that Vega shows surface structures has been reported by \cite{Bohm:2015}, who used high resolution spectroscopy to reveal line profile variability. The variability is compatible with rotational modulation of hot or cold 
starspots with a lifetime longer than about 5 days. Further inspection has shown a complex behavior, with a number of surface structures that appear stable, and others evolving on timescales of days \citep{2017MNRAS.472L..30P}. These starspots might be caused by the presence of small scale, weak magnetic fields at the stellar surface. If this correspondence is confirmed, the rapid evolution of the surface features would support a dynamo-generated origin of ultra-weak magnetism in A-stars. 

\section{Summary of Possible Sources of Magnetism in A Stars}\label{magnetism_summary}
Strong magnetic fields are found in a small fraction of A stars (the Ap phenomenon, which affects 5-10\% of A stars). These magnetic fields might be inherited during the star formation process, or generated in a  pre main sequence convective dynamo phase or during a stellar merger. Magnetic fields with amplitude between 300 and a few G are not detected \citep[the so-called ``magnetic desert''][]{2007AA4751053A}. The ultra-weak magnetic fields discussed in this paper, with surface amplitudes below a few G, might be common in A stars. We have proposed two possible scenarios for their origin, failed-fossils \citep{2013MNRAS.428.2789B} and dynamo-generated fields from the HeIICZ (this work). These two scenarios make different predictions for the amplitude and evolution of the magnetic field, as well as for the associated surface activity. 
Failed fossils should constantly decrease in amplitude as stars evolve on the main sequence, while during the same evolutionary phase the amplitude of dynamo generated fields is expected to increase (see e.g. Fig.~\ref{bsur_AB}). 
Moreover, dynamo-generated magnetic fields might have small-scales, rapidly-evolving features that are constantly regenerated, while in the fossil fields scenario small-scale features are expected to be removed quickly, leaving behind only the larger scales. 
This also means that stars with only failed fossil fields should be relatively inactive, while  stars hosting HeIICZ dynamo-generated fields could show some level of activity, although predicting exactly at which level is challenging (see Sec.~\ref{consequences}).

Another possibility is a hybrid failed-fossil-dynamo system. From the theoretical point of view it seems to be difficult to get completely rid of a weak large-scale field, that this field should not be strong enough to suppress subsurface convection, and that there should be some kind of convective dynamo producing a magnetic field. Unfortunately we don't know much about this dynamo and in particular whether and how it could produce large length scales. In the event that a thin-shell dynamo cannot produce the observed large length scales on its own, or even with the help of smearing out by differential rotation, a plausible solution would be for the dynamo to be given some large-scale structure or polarity-bias by an underlying failed-fossil field.

\section{Discussion}\label{disc}

\subsection{The effect of a fossil field}\label{fossil}

Some fraction of A stars, the Ap stars have large-scale, steady fields of $200$ G to $30$ kG \citep[see e.g.][for a review]{2009ASPC..405..473M}. These are fossil fields, anchored deep in the stellar interior. There is a one-to-one correlation \citep{2007AA4751053A} between these strong magnetic fields and chemical peculiarities at the surface caused by gravitational settling and radiative levitation, processes that are usually washed out by convection and surface turbulence in other A stars. This strongly suggests that a magnetic field can suppress at least some of the convection. 

A sufficiently strong field does indeed suppress convection,  as in sunspots, forcing an increase in temperature gradient so that the entire energy flux is transported radiatively. However, it is not obvious where the field strength threshold should be. One might na\"{i}vely expect convection to be suppressed by a field of greater energy density than the kinetic energy in the convective motion. However the work of \citet{1966MNRAS.133...85G},\citet{1969MNRAS.145..217M} and \citet{1970MSRSL..19..167M} suggests that in that case the temperature gradient simply steepens further above the adiabatic gradient until convection resumes, and that to suppress convection completely, a field at equipartition with the {\it thermal} energy, rather than the convection kinetic energy, is required. However, these studies consider a situation where the energy flux is entirely convective, such as is approximately the case in the bulk of the solar convective zone. In ABO-star subsurface convective layers, the situation is different in that only a small fraction ($\sim5\%$ or less) of the stellar heat flux is carried by convection, and that the temperature gradient is already significantly above adiabatic. It may be then that the threshold in this context is intermediate between convective and thermal equipartition.

An important clue comes from recent observations. \citet{2014arXiv1409.0028S} looked at a sample of O stars with and without detected magnetic fields, finding that whilst one star with a 20 kG field (NGC 1624-2) lacks measurable macroturbulence, the other stars in the sample, which have fields up to 2.5 kG, all display macroturbulent velocities of at least 20 km s$^{-1}$. This result is consistent with the threshold for suppression of convection being equipartition with thermal energy, and {\it probably} incompatible with the threshold being equipartition with the convective energy, since equipartition field strengths for the stars in this sample are about 1-2 kG \citep{2011A&A...534A.140C}. A larger sample including stars with fields between 2.5 and 20 kG should shed more light on the situation in the future.

In A stars the field strength thresholds for suppression of convection can simply be taken from the lower plate of Fig.~\ref{convective_efficiency}, equipartition with either convection or pressure. In the convective equipartition hypothesis, all subsurface convection should be suppressed in Ap stars. If however equipartition with thermal pressure is required, the threshold for HeII convection suppression is around $3$ kG, so that convection would be present in the weaker-field Ap stars, and even HeI convection may be present in less massive Ap stars at the bottom of the field-strength distribution. This suppression would have consequences on observables connected to this convection, e.g.\ microturbulence, and chemical abundances, as mentioned above.

Apart from Ap/Bp stars it is likely that the rest of the A and late-B population harbour weaker large-scale fields in their interiors. A convective dynamo during the pre-main sequence will leave behind a magnetic field which decays on the main sequence on a dynamic timescale given in terms of the Alfv\'en timescale and the star's rotation by $\tau_{\rm A}^2\Omega$, which goes to infinity as the field strength goes to zero. The field can therefore never disappear completely because the weaker it gets, the more slowly it decays. In \citet{2013MNRAS.428.2789B} we predicted, by equating this decay timescale to the stellar age, internal field strengths in Vega and Sirius of around 20 and 7 gauss respectively, with somewhat weaker fields at the surface. Unable to suppress convection, these fields would co-exist with it. That a large-scale field could be screened somehow below the convective layer, as suggested for instance by \citet{1998Natur.394..755G}, seems impossible \citep{2017RSOS....460271B}.

\subsection{Dependency on $\alphamlt$}\label{alphadependency}
\begin{figure}[h!]
\begin{center}
\includegraphics[width=1.00\columnwidth]{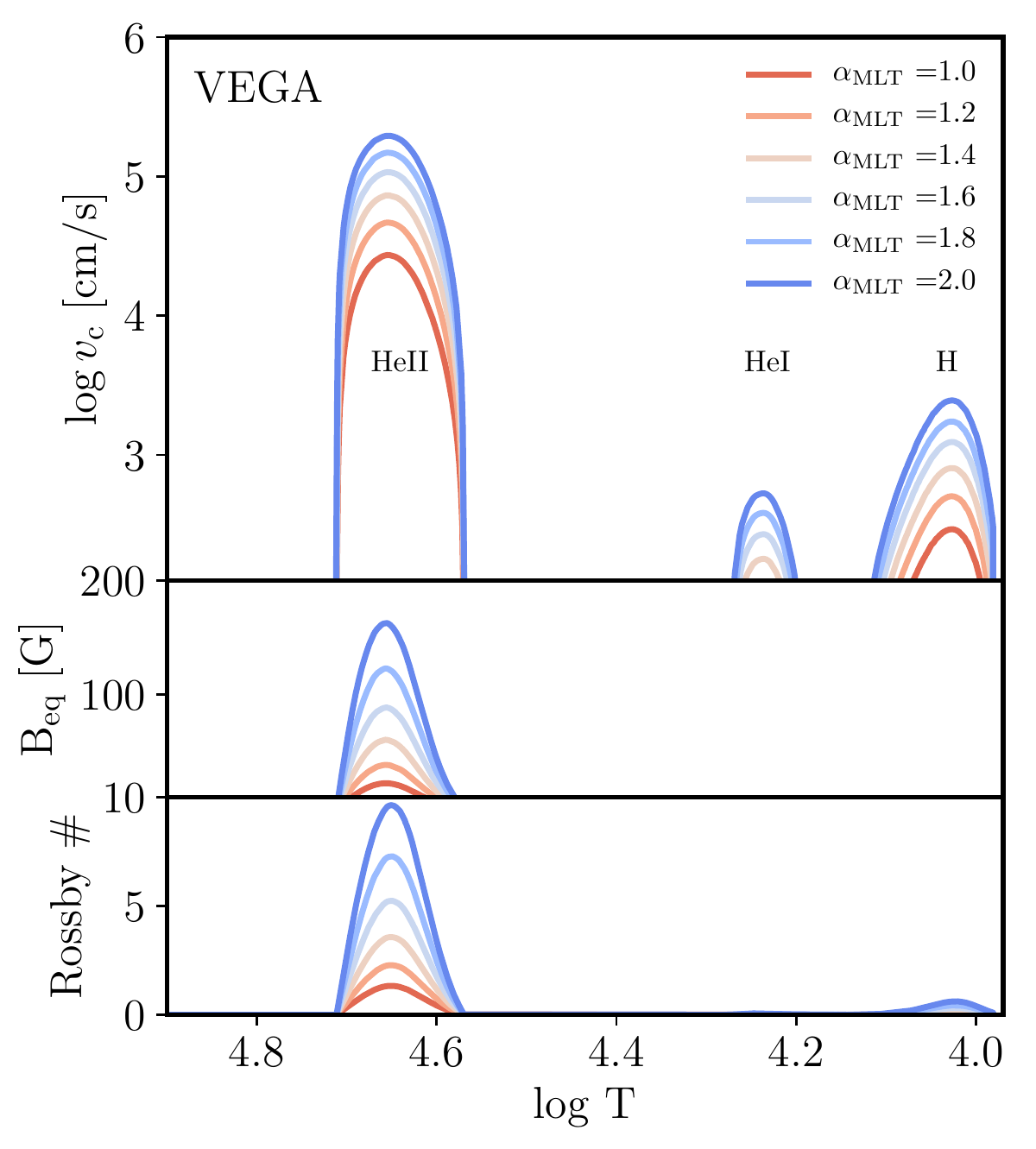}
\caption{{\label{Vegaconv_alpha} Top: Convective velocities as function of temperature in the outer layers of a $2.4\mso$ model with T$_{\rm eff}=9560$ K representative of the star Vega. The stellar surface is to the right. The velocities are in $\log$ (cm/s) and have been calculated using different values of the mixing length parameter $\alpha_{\rm MLT}$. Middle and Bottom: dependency of the equipartition magnetic field and Rossby number in the outer layers of the model as function of $\alpha_{\rm MLT}$. The Rossby number is calculated assuming a rotation period of 13 hours, appropriate for the rapidly rotating star Vega.%
}}
\end{center}
\end{figure}

In the case of efficient convection the velocities calculated by the Mixing Length Theory have a weak dependence on the unknown mixing length parameter $\vca \propto \alphamlt^{1/3}$ (see Appendix~\ref{MLT}). In the case of the (sub)surface convection zones studied here, convection is very inefficient. The small energy excess content carried by the convective element is lost before it can be advected because the thermal timescale of the convective element is comparable to the dynamical timescale of the convective motion. This is quantified by the Peclet number, which measures the ratio between the thermal and the dynamical timescale \citep[see e.g.][]{Maeder:2009}. The convective zones studied in this work have all small Peclet numbers. In the case of inefficient (non-adiabatic) convection, the MLT has to solve a cubic equation and the resulting dependency of the convective velocities on the uncertain  $\alphamlt$ parameter is much steeper, $\vca\propto\alphamlt^{3}$ (Appendix~\ref{MLT}). It is difficult to say what exact value of the $\alphamlt$ parameter is needed to reproduce the properties of stellar convection in these regions. When compared to 2D and 3D numerical simulations, a range of values between 1 and 2 are usually discussed \citep[see e.g.][]{Kupka:2017}, so that the values of velocity and resulting equipartition magnetic fields calculated in this work should be considered only as order of magnitude estimates. In Fig.~\ref{Vegaconv_alpha} we show how some of the relevant properties in our Vega model depend on the choice of the $\alphamlt$ parameters.

\section{Conclusions}\label{conc}
A complex landscape of surface and subsurface convective regions is present in the  outer 2\% in radius of A and late B stars.
These convection zones are driven by partial ionization of H and He and  have thickness comparable to the local pressure scale-height. Being so thin and close to the stellar surface, they are very inefficient, with a tiny amount of flux transported by convection. For temperatures above $\approx$ 8,000 K, the most efficient convection zone is due to second ionization of He. 

In the HeIICZ, convective velocities calculated using the Mixing Length Theory with $\alphamlt = 1.6$ result in equipartition magnetic fields of order 100 G for A Stars. For inefficient convection, the velocities calculated in the context of the MLT depend on the $\alphamlt$ parameter as $\vca \propto \alphamlt^{3}$. Since the value of the $\alphamlt$ is uncertain in these regions, the values of velocity and resulting equipartition magnetic fields should be considered only as order of magnitude estimates. 

We showed that, among advection, magnetic diffusion and magnetic buoyancy, the most likely process that can bring to surface magnetic fields generated by dynamo action in the subsurface convective zones of A stars is magnetic buoyancy. Interestingly, this process leads to the appearance of surface magnetic fields of amplitude comparable to the fields observed in Vega and in other A Stars with ultra-weak magnetic fields (1...10 G).

These magnetic fields can cause spots, which thanks to rotational modulation can lead to observable photometric variability in A and late B-type stars. 
These magnetic spots are expected to be bright because the surface is either radiative (for temperatures above 10,000-11,000 K) or convective, but with convection transporting a negligible amount of flux.

The spot temperature contrast is likely small, on the order of 10 K , which translates into less than 1\% local intensity fluctuations and a much lower integrated intensity fluctuation, depending on the spots filling factor. The occurrence and detectability of this type of spots decrease moving from A to late B-type stars. We predict regions of the \hrd where (sub)surface convection is unlikely to have any effect, and weak magnetic fields and photometric variability due to magnetic spots should be absent/undetectable in the majority of non Ap/Bp stars. At high luminosities, the appearance of the FeCZ is predicted to lead again to observable effects in early-type stars above $\log L \sim 10^{3.2}$ \citep[at solar metallicity, see ][]{2009A&A...499..279C,2011A&A...534A.140C}. 

It is not yet clear what the geometry of the field and resulting magnetic spots might look like, as this is related to the type of dynamo at work in the HeIICZ, which we do not investigate here. This said, recent observations of Vega seem to show a  relatively complex magnetic field, as well as surface structures evolving on fairly rapidly timescales. Further theoretical investigations are required to check if a dynamo in the tiny convective envelope regions of A stars can reproduce these observations.

The scenario described in this paper represents an alternative  to \citet{2013MNRAS.428.2789B} for explaining the presence of ultra-weak magnetic fields in A stars. Interestingly, the failed fossil field hypothesis described in our previous work makes different predictions than the dynamo-generated field scenario discussed in this paper. Dynamo-generated magnetic fields are expected to have small-scale, rapidly-evolving features, contrary to slowly-evolving, larger scale failed fossils. Moreover, the amplitude of surface magnetic fields as function of stellar age is expected to decrease in the failed fossil scenario, and increase in the dynamo-generated field scenario. Therefore, the rapidly increasing number of detections and observations of ultra-weak magnetic fields in intermediate stars, should allow to determine which scenario, if any, is to favor for the origin of this intriguing type of stellar magnetism.

\acknowledgments
MC thanks Alfio Bonanno, Fabio Del Sordo, Gustavo Guerrero, Pablo Marchant, Keaton Burns and Daniel Lecoanet for some very useful conversations on the topic of this paper. The Center for Computational Astrophysics at the Flatiron Institute is supported by the Simons Foundation.

\appendix
\section{Dependency of Convective Velocities on the $\alphamlt$ Parameter}\label{MLT}

Following e.g. \citet{CoxGiuli}, the three basic equations to be solved involve three unknowns: The ambient temperature gradient $\nabla$, the temperature gradient within a convective element $\nabla'$ and the convective efficiency $\Gamma$:
\begin{align}
\Gamma = A  \, (\nabla - \nabla')^{1/2} \label{1a} \\ 
\nabla_{r} - \nabla = a_0 \, A  \, (\nabla - \nabla')^{3/2} \label{1b}\\ 
\Gamma = (\nabla - \nabla')/ (\nabla' - \gradad). \label{1c}
\end{align}
Here $A$ is essentially the ratio of the convective  to the radiative conductivities
\begin{equation}\label{Aterm}
A \equiv \frac{Q^{1/2} \cp \kappa g \rho^{5/2} \alphamlt^2}{12 \sqrt{2} ac P^{1/2}T^3} ,
\end{equation}
where
\begin{equation}
Q = \frac{4-3 \beta}{\beta} - \bigg(\frac{\partial \ln \mu}{\partial \ln T} \bigg)_{P},
\end{equation}
with $\beta = P_{Gas}/P$.
The numerical factor $a_0$ in \ref{1b} is of order 1 and differs slightly depending on different implementation of the MLT. If convection is inefficient, $\nabla \approx \nabla'\approx \nabla_{r}$ and one can rewrite \ref{1c} as 
\begin{equation}\label{gamma}
\Gamma\simeq (\nabla - \nabla')/(\nabla_{r}-\gradad).    
\end{equation}
If we combine this equation for $\Gamma$ with \ref{1a} and \ref{Aterm} we obtain 
\begin{equation}
    (\nabla - \nabla')^{1/2} = A (\nabla_{r}-\gradad) \propto \alphamlt^2
\end{equation}
and since $v_{\textrm c} \propto \alphamlt (\nabla - \nabla')^{1/2}$, we find that $v_{\textrm c} \propto \alphamlt^3$. 

Note instead that for efficient convection, $\nabla \approx \nabla'\approx \gradad$ and using  \ref{1b} we find 
\begin{equation}
    (\nabla - \nabla')^{1/2} = \left(\frac{\nabla_{r}-\gradad}{a_0 A}\right)^{1/3} \propto \alphamlt^{-2/3} 
\end{equation}
and $v_{\textrm c} \propto \alphamlt^{1/3}$.

\bibliography{biblio.bib}
\end{document}